\documentclass[a4paper,10.9pt]{article}
\usepackage{bbding}
\usepackage{}
\usepackage{amsmath}
\usepackage{mathrsfs}
\usepackage{graphicx, subfigure}
\usepackage{wasysym}
\usepackage{color}
\usepackage{bm}
\usepackage{pifont}
\usepackage{txfonts}
\usepackage{graphicx}
\usepackage{amsfonts}
\usepackage{amssymb}
\usepackage{mathrsfs,psfrag,eepic,epsfig}
\makeatletter \@addtoreset{equation}{section}

\makeatletter
\newfont{\footsc}{cmcsc10 at 8truept}
\newfont{\footbf}{cmbx10 at 8truept}
\newfont{\footrm}{cmr10 at 10truept}
\title{\bf{The nonlinear steepest descent approach for long time behavior of the two-component coupled Sasa-Satsuma equation with a $5\times 5$ Lax pair} \footnote{This work is supported by the National Key Research and Development Program of
China under Grant No. 2017YFB0202901 and the National Natural Science Foundation of China under Grant No.11871180.\protect\\
\hspace*{3ex} $^{*}$Corresponding authors.\protect\\
\hspace*{3ex} E-mail address: xiubinwang@163.com (X.B. Wang) and bohan@hit.edu.cn (B. Han)}}

\author{Xiu-Bin Wang$^{*}$ and Bo Han$^{*}$ \\
\small Department of Mathematics, Harbin Institute of Technology, Harbin 150001,
People's Republic of China}

\date{}
\begin{document}
\maketitle

\noindent{\large \bf Abstract:}
Under investigation in this work is the coupled Sasa-Satsuma equation,
which can describe the propagations of two optical pulse envelopes in birefringent fibers.
The Riemann-Hilbert problem for the equation is formulated on the basis of the corresponding $5\times5$ matrix spectral problem,
which allows us to present a suitable representation for the solution of the equation.
Then the Deift-Zhou steepest descent method is used to analyze the long time behavior of the coupled Sasa-Satsuma equation.
\\
%{\bf PACS numbers:} 02.30.Jr, 02.30.Ik, 05.45.Yv.\\

%\begin{center}
%(Some figures in this article are in colour only in the electronic version)
%\end{center}

\section{Introduction}

As we all know, the nonlinear Schr\"{o}dinger (NLS) equation is a key integrable system in the field of mathematical physics.
There are many physical phenomenon where the NLS equation appears.
For instance, the NLS equation can describe slowly varying wave envelopes in dispersive media from water waves, nonlinear optics, and plasma physics.
In particular, the NLS equation can also describe the soliton propagation in optical fibers where only the group
velocity dispersion and the self-phase modulation effects are discussed.
However, for ultrashort pulse in optical fibers, the effects of the self steepening, the third-order dispersion,
and the stimulated Raman scattering should be taken into account. Because of these effects,
the dynamic behaviors of the ultrashort pulses can be described by the higher-order NLS equation (also called Sasa-Satsuma equation) \cite{SS-1991}-\cite{JJC-2015}
\begin{equation}\label{SS-1}
q_{T}+\frac{1}{2}q_{XX}+i\left\{q_{1XXX}+6|q|^{2}q_{X}+3q\left(|q|^{2}\right)_{X}\right\}=0,\\
\end{equation}
where $q=q(X,T)$ is a complex-valued function.
In addition,
to model the propagations of two optical pulse envelopes
in birefringent fibers well, some coupled Sasa-Satsuma equations were also proposed and discussed \cite{KN-1998}-\cite{AM-2002}.
In this work, we therefore focus on a coupled Sasa-Satsuma (CSS in brief) equation
\begin{equation}\label{CSS}
\left\{ \begin{aligned}
  iq_{1T}+\frac{1}{2}q_{1XX}&+\left(|q_{1}|^{2}+|q_{2}|^{2}\right)q_{1}+i\left\{q_{1XXX}+6\left(|q_{1}|^{2}
  +q_{2}|^{2}\right)q_{1X}+3q_{1}\left(|q_{1}|^{2}\right.\right.\\
  &\left.\left.+q_{2}|^{2}\right)_{x}\right\}=0,\\
  iq_{2T}+\frac{1}{2}q_{2XX}&+\left(|q_{1}|^{2}+|q_{2}|^{2}\right)q_{2}+i\left\{q_{2XXX}+6\left(|q_{1}|^{2}
  +q_{2}|^{2}\right)q_{2X}+3q_{2}\left(|q_{1}|^{2}\right.\right.\\
  &\left.\left.+q_{2}|^{2}\right)_{x}\right\}=0,
                     \end{aligned} \right.
\end{equation}
which can be rewritten in the following form \cite{KN-1998}
\begin{equation}\label{SS}
\left\{ \begin{aligned}
  &u_{t}+\left\{u_{xxx}+6\left(|u|^{2}+|v|^{2}\right)u_{x}+3u\left(|u|^{2}+|v|^{2}\right)_{x}\right\}=0,\\
  &v_{t}+\left\{v_{xxx}+6\left(|u|^{2}+|v|^{2}\right)v_{x}+3v\left(|u|^{2}+|v|^{2}\right)_{x}\right\}=0,\\
   &u(x,0)=u_{0}(x),~~v(x,0)=v_{0}(x),
                     \end{aligned} \right.
\end{equation}
by introducing the gauge, Galilean and scale transformations
\begin{equation*}\label{SS-1}
\left\{ \begin{aligned}
  &u(x,t)=q_{1}(X,T)\exp\left[-\frac{i}{6}\left(X-\frac{T}{18}\right)\right],\\
  &v(x,t)=q_{2}(X,T)\exp\left[-\frac{i}{6}\left(X-\frac{T}{18}\right)\right],\\
  &x=X-\frac{T}{12\epsilon},~~t=T,
                       \end{aligned} \right.
\end{equation*}
where $(u_{0},v_{0})$ lie in the Schwartz space.
Besides, $q_{1}=q_{1}(X,T)$ and $q_{2}=q_{2}(X,T)$ are two complex functions of variables $X,T$.
The CSS equation \eqref{SS} is still completely integrable.
Additionally, the CSS equation \eqref{SS} has also been investigated
via Darboux transformation, Darboux-B\"{a}cklund transformation and Hirota method etc.
Recently, we have also studied the long-time behavior and rogue wave solutions of the integrable
three-component coupled nonlinear Schr\"{o}dinger equation \cite{wxb-2018,wxb-2019}.
Recently, one of our authors Wang have provided the characteristics of the breather
and rogue waves in a (2+1)-dimensional NLS equation \cite{wxb-ams}.
In this paper, we will consider the long-time asymptotics
of the CSS equation \eqref{SS} on the line.

More recent years,
there are many investigations on long time asymptotics and exact solutions of nonlinear evolution equations \cite{AS-1997}-\cite{AS-2005}.
It is also known that the Deift-Zhou steepest descent approach is a powerful skill to
analyze long time behavior \cite{PD-1993}-\cite{WDS-2018}.
However, since Eq.\eqref{SS} contains a $5\times5$ matrix spectral problem,
the long time asymptotics for Eq.\eqref{SS} is rather complicated to consider.
The research in this direction, to the best of our knowledge, has not been conducted before.
The main objective of the present article is to analyze the long time asymptotics of Eq.\eqref{SS}
by utilizing the Riemann-Hilbert problem (RHP) via the Deift-Zhou steepest descent method.

The structure of this paper is given as follows.
In section, we derive a $5\times 5$ matrix RHP and find that the solution of Eq.\eqref{SS}
can be given by the solution of this RHP.
In Section 3, we obtain the main conclusion of this work by using the Deift-Zhou steepest descent method.
Finally, the further discussions are provided in section 4.

\section{Riemann-Hilbert Problem}
%In the present section,  we aim to formulate a RH problem to solve the Cauchy problem of the CSS equation \eqref{SS} with $\epsilon=1$.

System \eqref{SS} is still completely integrable. Its Lax pair yields \cite{gxg-2017}
\begin{equation}\label{SA-1}
\left\{ \begin{aligned}
  &\psi_{x}(x,t;\lambda)=i\lambda\sigma\psi(x,t;\lambda)+\textbf{U}(x,t)\psi(x,t;\lambda),\\
  &\psi_{t}(x,t;\lambda)=4i\lambda^{3}\sigma\psi(x,t;\lambda)+\textbf{V}(x,t;\lambda)\psi(x,t;\lambda),
                        \end{aligned} \right.
\end{equation}
where
\begin{equation}\label{SA-3}
\sigma=\left(
         \begin{array}{cc}
           \mathcal {I}_{4\times 4} & \textbf{0} \\
           \textbf{0} & -1 \\
         \end{array}
       \right),~~\textbf{U}=\left(
                              \begin{array}{cc}
                                \textbf{0}_{4\times 4} & \mathcal {U} \\
                                -\mathcal {U}^{\dag} & 0 \\
                              \end{array}
                            \right),~~
\mathcal {U}=\left(
    \begin{array}{c}
      \textbf{u} \\
      \textbf{v} \\
    \end{array}
  \right),~~\textbf{u}=\left(
                \begin{array}{c}
                  u \\
                  \bar{u} \\
                \end{array}
              \right),~~\textbf{v}=\left(
                            \begin{array}{c}
                              v \\
                              \bar{v} \\
                            \end{array}
                          \right),
\end{equation}
with
\begin{equation}\label{SA-4}
\textbf{V}(x,t,\lambda)=4\lambda^2\textbf{U}-2i\lambda\sigma\left(\textbf{U}_{x}-\textbf{U}^{2}\right)
+\left(\textbf{U}_{x}U-\textbf{U}\textbf{U}_{x}\right)-\textbf{U}_{xx}+2\textbf{U}^{3}.
\end{equation}
Here the overbar represents the complex conjugation and ``\dag'' represents Hermitian of a matrix.

In the following, by introducing a new matrix function
\begin{equation}\label{SA-6}
\psi(x,t;\lambda)=\mu(x,t;\lambda)e^{i\left(\lambda x+4i\lambda^3t\right)\sigma}.
\end{equation}
The spectral problem \eqref{SA-1} then gives
\begin{equation}\label{SA-7}
\left\{ \begin{aligned}
&\mu_{x}(x,t;\lambda)-i\lambda[\sigma,\mu(x,t;\lambda)]=\textbf{U}(x,t)\mu(x,t;\lambda),\\
&\mu_{t}(x,t;\lambda)-4i\lambda^3[\sigma,\mu(x,t;\lambda)]=\textbf{V}(x,t;\lambda)\mu(x,t;\lambda).
                       \end{aligned} \right.
\end{equation}

We next present two eigenfunctions $\mu_{\pm}(x,t;\lambda)$ of x-part of \eqref{SA-7} by the following Volterra type integral equations
\begin{equation}\label{SA-9}
\mu_{\pm}=\mathcal {I}+\int_{\pm\infty}^{x} e^{-i\lambda(x-\xi)\hat{\sigma}}[\textbf{U}(\xi,t)\mu_{\pm}(\xi,t;\lambda)]d\xi,
\end{equation}
where $\hat{\sigma}$ represents the operators which act on a $5\times 5$ matrix $\Omega$ by $\hat{\sigma}=[\sigma,\Omega]$.
Here $e^{\hat{\sigma}}=e^{\sigma}\Omega e^{\sigma}$.
Then we rewrite $\mu_{\pm}(x,t;\lambda)$ as
\begin{equation}\label{SSJ-1}
\mu_{\pm}(x,t;\lambda)=\left(\mu_{\pm L}(x,t;\lambda),\mu_{\pm R}(x,t;\lambda)\right),
\end{equation}
in which the first fourth columns of $\mu_{\pm}(x,t;\lambda)$ and fifth column are expressed by $\mu_{\pm L}(x,t;\lambda)$, respectively.
From \eqref{SA-9}, we can know that $\mu_{+L},\mu_{-R}$ and $\mu_{-L},\mu_{+R}$ are analytic in
$\mathbb{C}_{-}$ and $\mathbb{C}_{+}$, respectively. Furthermore
\begin{equation*}\label{SSJ-2}
\left\{ \begin{aligned}
&\left(\mu_{+L}(x,t;\lambda),\mu_{-R}(x,t;\lambda)\right)=\mathcal {I}+O\left(\frac{1}{\lambda}\right),~~\lambda\in\mathbb{C}_{-}\rightarrow\infty,\\
&\left(\mu_{-L}(x,t;\lambda),\mu_{+R}(x,t;\lambda)\right)=\mathcal {I}+O\left(\frac{1}{\lambda}\right),~~\lambda\in\mathbb{C}_{+}\rightarrow\infty,
                      \end{aligned} \right.
\end{equation*}

The solutions of the equation of differential equation \eqref{SA-7} can be related by a matrix independent of x and t; As a result
\begin{equation}\label{SA-12}
\mu_{-}(x,t;\lambda)=\mu_{+}(x,t;\lambda)e^{i\left(\lambda x+4i\lambda^3t\right)\hat{\sigma}}s(\lambda).
\end{equation}
Evaluation at $t=0$ arrives at
\begin{equation}\label{SA-13}
s(\lambda)=\lim_{x\rightarrow+\infty}e^{-i\lambda x\hat{\sigma}}\mu_{-}(x,0;\lambda),
\end{equation}
i.e.,
\begin{equation}\label{SA-13}
s(\lambda)=\mathcal {I}+\int_{-\infty}^{+\infty}e^{-i\lambda x\hat{\sigma}}\left[\textbf{U}(x,0)\mu_{-}(x,0;\lambda)\right]dx.
\end{equation}
The fact that $\mbox{tr}(U)=0$ together with Eq.\eqref{SA-9} indicates
\begin{equation}\label{SA-14-1}
\det(\mu_{\pm}(x,t;\lambda))=1.
\end{equation}
Therefore, one can obtain
\begin{equation}\label{SA-14}
 \det(s(\lambda))=1.
\end{equation}
Additionally, %if we define $U(x,t;\lambda)=-i\lambda\sigma+U(x,t)$, then
we can know that
\begin{equation}\label{SA-15}
\textbf{U}^{\dag}(x,t;\bar{\lambda})=-\textbf{U}(x,t;\lambda),~~\overline{\textbf{U}(x,t;-\bar{\lambda})}=\nabla\textbf{U}(x,t;\lambda)\nabla,
\end{equation}
where
\begin{equation*}\label{SA-16}
\nabla=\left(
         \begin{array}{ccccc}
           0 & 1 & 0 & 0 & 0 \\
           1 & 0 & 0 & 0 & 0 \\
           0 & 0 & 0 & 1 & 0 \\
           0 & 0 & 1 & 0 & 0 \\
           0 & 0 & 0 & 0 & 1 \\
         \end{array}
       \right).
\end{equation*}
Furthermore, it follows from \eqref{SA-1} that
\begin{equation}\label{SA-17}
\psi_{x}^{A}(x,t;\lambda)=\left(i\lambda\sigma-\textbf{U}(x,t)\right)^{T}\psi^{A}(x,t;\lambda),
\end{equation}
with $\psi^{A}(x,t;\lambda)=(\psi^{-1}(x,t;\lambda))^{T}$, where the superscript `T' represents a matrix transpose.
Consequently, we have
\begin{equation}\label{SA-181}
\psi^{\dag}(x,t;\bar{\lambda})=\psi^{-1}(x,t;\lambda),~~\psi(x,t;\lambda)=\nabla\overline{\psi(x,t;-\bar{\lambda})}\nabla.
\end{equation}
These relations indicate that the eigenfunctions $\mu_{j}(x,t;\lambda)$ meet
\begin{equation}\label{SA-18}
\mu^{\dag}(x,t;\bar{\lambda})=\mu^{-1}(x,t;\lambda),~~\mu(x,t;\lambda)=\nabla\overline{\mu(x,t;-\bar{\lambda})}\nabla,~~j=1,2,
\end{equation}
where `\dag' represents the Hermitian conjugate.
To sum up, the matrix-valued function $s(\lambda)$ admits the following symmetries
\begin{equation}\label{SA-19}
s^{\dag}(\bar{\lambda})=s^{-1}(\lambda),~~s(-\lambda)=\nabla \overline{s(\bar{\lambda})}\nabla.
\end{equation}
In the following, without otherwise specified, by matrix blocking we rewrite the $5\times5$ matrix \textbf{A} as
\begin{equation*}\label{RHP-22}
\textbf{A}=\left(
    \begin{array}{cc}
      \textbf{A}_{11} & \textbf{A}_{12} \\
      \textbf{A}_{21} & \textbf{A}_{22} \\
    \end{array}
  \right),
\end{equation*}
where $\textbf{A}_{11}$ is a $4\times 4$ matrix and $\textbf{A}_{22}$ is scalar. It follows from \eqref{SA-12}-\eqref{SA-19} that
\begin{align}\label{RHH-1}
&s_{22}^{\dag}(\bar{\lambda})=\det\left(s_{11}(\lambda)\right),~~s_{11}(\lambda)=\sigma_{1}\bar{s}_{11}(-\bar{\lambda})\sigma_{1},\notag\\
&s_{12}^{\dag}(\bar{\lambda})=-s_{21}\mbox{adj}(s_{11}(\lambda)),~~\bar{s}_{21}(-\bar{\lambda})\sigma_{1}=s_{21}(\lambda),
\end{align}
where
\begin{equation}\label{RHH-2}
\sigma_{1}=\left(
             \begin{array}{cccc}
               0 & 1 & 0 & 0 \\
               1 & 0 & 0 & 0 \\
               0 & 0 & 0 & 1 \\
               0 & 0 & 1 & 0 \\
             \end{array}
           \right),
\end{equation}
and $\mbox{adj}(\textbf{B})$ represents the adjoint matrix of matrix $\textbf{B}$.
Because of the above expression \eqref{RHH-2}, we can rewrite $s(\lambda)$ as
\begin{equation}\label{RHH-3}
s(\lambda)=\left(
             \begin{array}{cc}
               a(\lambda) & -\mbox{adj}(a^{\dag}(\bar{\lambda}))b^{\dag}(\bar{\lambda}) \\
               b(\lambda) & \det(a^{\dag}(\bar{\lambda})) \\
             \end{array}
           \right),
\end{equation}
where
\begin{equation}\label{RHH-4}
a(\lambda)=\sigma_{1}\bar{a}(-\bar{\lambda})\sigma_{1},~~\bar{b}(-\bar{\lambda})\sigma_{1}=b(\lambda).
\end{equation}
It follows that $a(\lambda)$ and $b(\lambda)$ satisfy
\begin{equation}\label{RHH-5}
\left\{ \begin{aligned}
&a(\lambda)=\mathcal {I}+\int_{-\infty}^{+\infty}\mathcal {U}(\xi,0)\mu_{-,21}(\xi,0;\lambda)dx,\\
&b(\lambda)=-\int_{-\infty}^{+\infty}e^{-2i\xi\lambda}\mathcal {U}^{\dag}(\xi,0)\mu_{-,11}(\xi,0;\lambda)d.
                       \end{aligned} \right.
\end{equation}
Obviously, $a(\lambda)$ is analytic in $\mathbb{C}_{+}$.

Suppose that $\det(a(\lambda))$ admits $4N$ simple zeros $\lambda_{1},\ldots, \lambda_{2N}$ in $\mathbb{C}_{+}$,
where $\lambda_{N+j}=-\bar{\lambda}_{j},j=1,2,\ldots 2N$. Define
\begin{equation}\label{RHH-6}
M(\xi;\lambda)=
\left\{ \begin{aligned}
&\left(\mu_{-L}(\lambda)a^{-1}(\lambda),\mu_{+R}(\lambda)\right),~~~~~~~\lambda\in\mathbb{C}_{+},\\
&\left(\mu_{+L}(\lambda),\mu_{-R}(\lambda)/\det a^{\dag}(\bar{\lambda})\right),~~\lambda\in\mathbb{C}_{-}.
                       \end{aligned} \right.
\end{equation}

\noindent
\textbf{Theorem 1.}
Let $a(\lambda)$ and $b(\lambda)$ be determined by \eqref{RHH-5}. Then
$M(x,t;\lambda)$ given by \eqref{RHH-6} satisfies the following matrix RHP.
We find a meromorphic function $M(x,t;\lambda)$ with simple poles at $\{\lambda_{j}\}_{1}^{4N}$ and $\{\bar{\lambda}_{j}\}_{1}^{4N}$,
Then it admits
\begin{equation}\label{RHH-7}
\left\{ \begin{aligned}
&M_{+}(\lambda)=M_{-}(\lambda)J(\lambda),~~\lambda\in\mathbb{R},\\
&M(\lambda)=\mathcal {I}+O\left(\frac{1}{\lambda}\right),~~~\lambda\rightarrow\infty,
                       \end{aligned} \right.
\end{equation}
and residue conditions
\begin{equation}\label{RHH-8}
\left\{ \begin{aligned}
&\mbox{Res}_{\lambda_{j}}M(\lambda)
=\lim_{\lambda\rightarrow\lambda_{j}}M(\lambda)\left(
                                                                                   \begin{array}{cc}
                                                                                     0 & 0 \\
                                                                                   \frac{  e^{-2i\theta(\lambda)t}b(\lambda)\mbox{adj}(a(\lambda))}{\dot{\det}(a(\lambda))} & 0 \\
                                                                                   \end{array}
                                                                                 \right),\\
&\mbox{Res}_{\bar{\lambda}_{j}}M(\lambda)
=\lim_{\lambda\rightarrow\bar{\lambda}_{j}}M(\lambda)\left(
\begin{array}{cc}
0 & -\frac{ e^{2i\theta(\lambda)t}\mbox{adj}(a^{\dag}(\bar{\lambda}))b(\bar{\lambda})^{\dag}}{\dot{\det}(a^{\dag}(\bar{\lambda}))} \\
0 & 0 \\
\end{array}
\right),
                       \end{aligned} \right.
\end{equation}
where $j=1,2,\ldots,2N$, $\dot{f}(\lambda)=df(\lambda)/d\lambda$,
\begin{align}\label{RHH-9}
&M_{\pm}=\lim_{\epsilon\rightarrow 0^{+}}M(\lambda\pm i\epsilon),~~
\gamma(\lambda)=b(\lambda)a^{-1}(\lambda),~~\lambda\in\mathbb{R},\notag\\
&J(\lambda)=\left(
             \begin{array}{cc}
               \mathcal {I}+\gamma^{\dag}(\bar{\lambda})\gamma(\lambda) & e^{2i\theta t}\gamma^{\dag}(\bar{\lambda})\\
               e^{-2i\theta t}\gamma(\lambda) & 1 \\
             \end{array}
           \right),~~\theta=\lambda\left(\frac{x}{t}+4\lambda^2\right).
\end{align}
Here $\gamma(\lambda)$ lies in Schwartz space and satisfies
\begin{equation*}\label{RHH-10}
\gamma(\lambda)=\gamma^{\dag}(-\bar{\lambda})\sigma_{1},~~\sup_{\lambda\in\mathbb{R}}\gamma(\lambda)<\infty.
\end{equation*}
Let
\begin{equation}\label{RHP-26}
\mathcal {U}(x,t)=\left(
         \begin{array}{c}
           \textbf{u} \\
           \textbf{v} \\
         \end{array}
       \right)
=2i\lim_{\lambda\rightarrow\infty}\left(\lambda M(x,t;\lambda)\right)_{12}.
\end{equation}
Then $u(x,t)$, $v(x,t)$ can represent the solution of the CSS equation \eqref{SS}.

\section{Long-time asymptotic analysis}
According to the idea of Deift and Zhou \cite{PD-1993},
we next consider the stationary points of the function $\theta$, i.e., setting
$\frac{d\theta}{d\lambda}=0$,
the stationary phase points are constructed for $x>0$ as
$\pm\lambda_{0}=\pm\sqrt{\frac{x}{12t}}$,
Thus,
$\theta=4\lambda\left(\lambda^2-3\lambda_{0}^2\right)$.
In what follows, we mainly focus on physically interesting region $\lambda_{0}\in(0,C]$,
where $C$ is a constant.

$~~~~~~~~~~~~~~~~~~~~~~~~~~~$
{\rotatebox{0}{\includegraphics[width=8.0cm,height=6.5cm,angle=0]{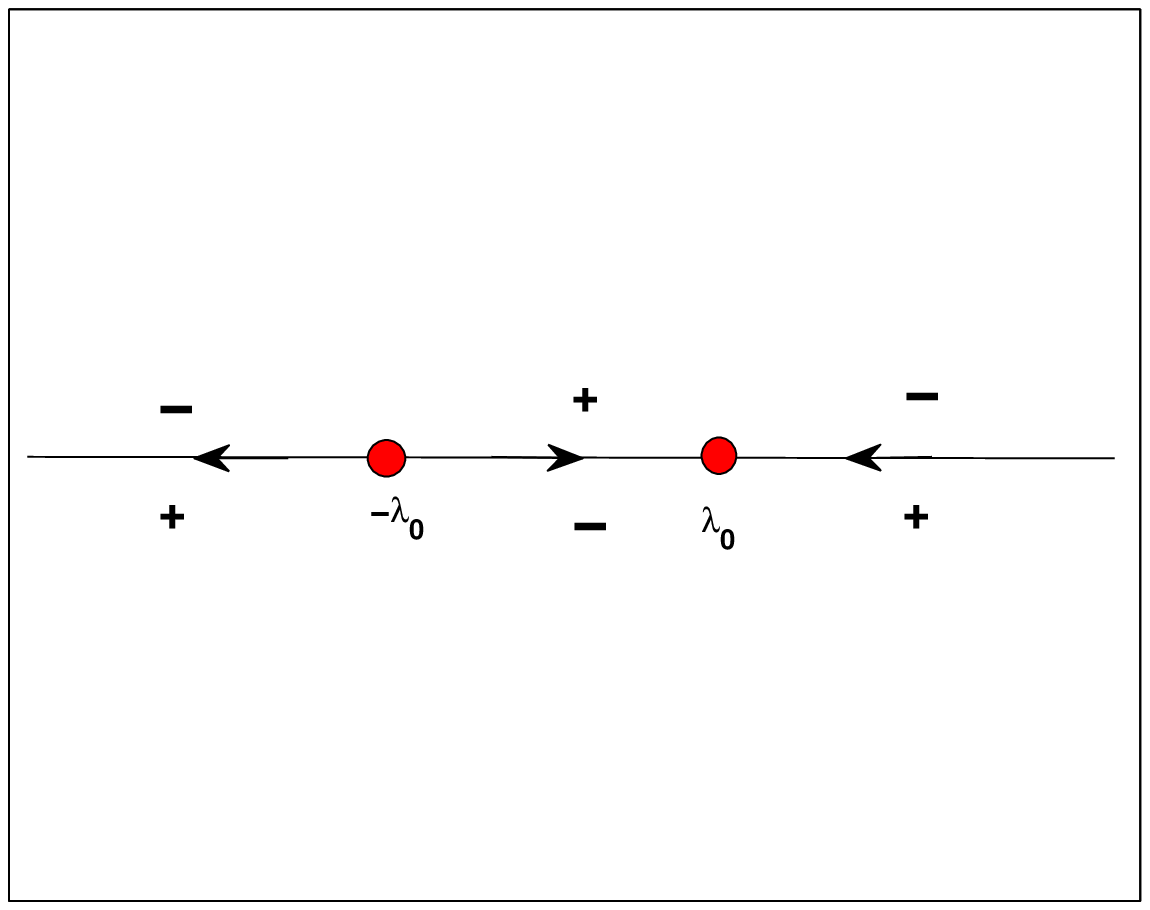}}}\\
\noindent {\small \textbf{Figure 1.} The oriented jump contour $\mathbb{R}$. \\}

\subsection{Factorization of the jump matrix}
We notice that the jump matrix admits two distinct factorizations
\begin{equation}\label{TRH-1}
J=\left\{ \begin{aligned}
&\left(
    \begin{array}{cc}
      \mathcal {I} & e^{2i\theta}\gamma^{\dag}(\bar{\lambda}) \\
      0 & 1 \\
    \end{array}
  \right)\left(
           \begin{array}{cc}
             \mathcal {I} & 0 \\
             e^{-2i\theta}\gamma(\lambda) & 1 \\
           \end{array}
         \right),\\
&\left(
  \begin{array}{cc}
    \mathcal {I} & 0 \\
    \frac{e^{-2i\theta}\gamma(\lambda)}{1+\gamma(\lambda)\gamma^{\dag}(\bar{\lambda})} & 1 \\
  \end{array}
\right)\left(
         \begin{array}{cc}
           \left(\mathcal {I}+\gamma^{\dag}(\bar{\lambda})\gamma(\lambda)\right) & 0 \\
           0 & \frac{1}{1+\gamma(\lambda)\gamma^{\dag}(\bar{\lambda})} \\
         \end{array}
       \right)\left(
                \begin{array}{cc}
                  \mathcal {I} & \frac{e^{2it\theta\gamma^{\dag}}(\bar{\lambda})}{1+\gamma(\lambda)\gamma^{\dag}(\bar{\lambda})} \\
                  0 & 1 \\
                \end{array}
              \right).
                                \end{aligned} \right.
\end{equation}
We next consider a function $\delta(\lambda)$ as the solution of the matrix problem
\begin{equation}\label{TRH-2}
\left\{ \begin{aligned}
&\delta_{+}(\lambda)=\left(\mathcal {I}+\gamma^{\dag}\gamma\right)\delta_{-}(\lambda),~~|\lambda|<\lambda_{0},\\
&~~~~~~~~=\delta_{-}(\lambda),~~~~~~~~~~~~~~~~~~~|\lambda|>\lambda_{0},\\
&\delta(\lambda)\rightarrow \mathcal {I},~~~~~~~~~~~~~~~~~~~~~~~~~~\lambda\rightarrow\infty.
                                \end{aligned} \right.
\end{equation}
As the jump matrix $\left(\mathcal {I}+\gamma^{\dag}\gamma\right)$ is positive definite,
the vanishing lemma gives the existence and uniqueness of the function $\delta(\lambda)$.
Moreover, we have
\begin{equation*}\label{TRH-3}
\left\{ \begin{aligned}
&\det(\delta_{+}(\lambda))=\left(1+|\gamma|^{2}\right)\det(\delta_{-}(\lambda)),~~|\lambda|<\lambda_{0},\\
&~~~~~~~~=\det(\delta_{-}(\lambda)),~~~~~~~~~~~~~~~~~~~~~~~~~~|\lambda|>\lambda_{0},\\
&\det(\delta(\lambda))\rightarrow 1,~~~~~~~~~~~~~~~~~~~~~~~~~~~~~~~~\lambda\rightarrow\infty.
                                \end{aligned} \right.
\end{equation*}
By utilizing the Plemelj formula \cite{MJ-2003}, we can get
\begin{equation}\label{TRH-4}
\det(\delta(\lambda))=\exp\left\{\frac{1}{2\pi i}\int_{-\lambda_{0}}^{\lambda_{0}}\frac{\log(1+\gamma(\xi)\gamma^{\dag}(\xi))}{\xi-\lambda}d\xi\right\}
=\left(\frac{\lambda+\lambda_{0}}{\lambda-\lambda_{0}}\right)^{i\nu}e^{\chi(\lambda)},
\end{equation}
where
\begin{equation*}\label{TRH-5}
\left\{ \begin{aligned}
&\nu=\frac{1}{2\pi}\log\left(1+\gamma(\lambda_{0})\gamma^{\dag}(\lambda_{0})\right)>0,\\
&\chi(\lambda)=\frac{1}{2\pi i}\int_{-\lambda_{0}}^{\lambda_{0}}
\log\left(\frac{1+\gamma(\xi)\gamma^{\dag}(\xi)}{1+\gamma(\lambda_{0})\gamma^{\dag}(\lambda_{0})}\right)
\frac{d\xi}{\xi-\lambda}.
      \end{aligned} \right.
\end{equation*}
Then we have used the following relation
\begin{equation}\label{TRH-6}
1+\gamma(\lambda_{0})\gamma^{\dag}(\lambda_{0})=1+\gamma(-\lambda_{0})\gamma^{\dag}(-\lambda_{0}),
\end{equation}
which can be obtained from the second symmetry condition in \eqref{SA-9}.

In addition, for $|\lambda|<\lambda_{0}$, it follows from \eqref{TRH-2} that
\begin{equation}\label{TRH-7}
\lim_{\epsilon\rightarrow0^{+}}\delta(\lambda- i\epsilon)=\left(\mathcal {I}+\gamma(\lambda)^{\dag}\gamma(\lambda)\right)^{-1}
\lim_{\epsilon\rightarrow0^{-}}\delta(\lambda+ i\epsilon).
\end{equation}
If we set $g(\lambda)=\left(\delta^{\dag}(\bar{\lambda})\right)^{-1}$, then we can get
\begin{equation}\label{TRH-8}
g_{+}(\lambda)=\left(\mathcal {I}+\gamma^{\dag}(\lambda)\gamma(\lambda)\right)g_{-}(\lambda).
\end{equation}
Thus, we know
\begin{equation}\label{TRH-8}
\left(\delta^{\dag}(\bar{\lambda})\right)^{-1}=\delta(\lambda).
\end{equation}
Similar to \cite{PD-1993}, after a direct calculation, we can obtain
\begin{equation}\label{TRH-9}
|\delta(\lambda)|\leq\mbox{const}<\infty,~~|\det\left(\delta(\lambda)\right)|\leq\mbox{const}<\infty,
\end{equation}
for all $\lambda$, where we define $|\textbf{A}|=\sqrt{\left(\mbox{tr}\textbf{A}^{\dag}\textbf{A}\right)}$ for any matrix \textbf{A}.
The define
\begin{equation*}\label{TRH-10}
\Delta(\lambda)=\left(
                  \begin{array}{cc}
                    \delta(\lambda)^{-1} & 0 \\
                    0 & \det\left(\delta(\lambda)\right) \\
                  \end{array}
                \right).
\end{equation*}
Introduce
\begin{equation}\label{TRH-11}
M^{\Delta}(x,t;\lambda)=M(x,t;\lambda)\Delta(\lambda),
\end{equation}
and reverse the orientation for $|\lambda|<\lambda_{0}$ as seen in Fig.1.

\noindent
\textbf{Theorem 2.}
The $M^{\Delta}$ admits the following RHP
\begin{equation}\label{TRH-12}
\left\{ \begin{aligned}
&M^{\Delta}_{+}(x,t;\lambda)=M_{-}^{\Delta}(x,t;\lambda)J^{\Delta}(x,t;\lambda),~~\lambda\in\mathbb{R},\\
&M^{\Delta}(x,t;\lambda)\rightarrow\mathcal {I},~~~~~~~~~~~~\lambda\rightarrow\infty,
      \end{aligned} \right.
\end{equation}
where
\begin{align}\label{TRH-13}
&J^{\Delta}(\lambda)=\left(
              \begin{array}{cc}
                \mathcal {I} & 0 \\
                \frac{e^{-2it\theta}\rho^{\dag}(\bar{\lambda})\delta^{-1}_{-}(\lambda)}{\det \delta_{-}(\lambda)} & 1 \\
              \end{array}
            \right)\left(
                     \begin{array}{cc}
                       \mathcal {I} & \left(\det\delta_{+}(\lambda)\right)e^{2it\theta}\delta_{+}(\lambda)\rho(\lambda) \\
                       0 & 1 \\
                     \end{array}
                   \right),
\end{align}
and the vector-valued function
\begin{equation*}\label{TRH-14}
\rho(\lambda)=\left\{ \begin{aligned}
&\frac{\gamma^{\dag}(\bar{\lambda})}{1+\gamma(\lambda)\gamma^{\dag}(\bar{\lambda})},~~|\lambda|<\lambda_{0},\\
&-\gamma^{\dag}(\bar{\lambda}),~~~~~~~~~~~~|\lambda|>\lambda_{0}.
      \end{aligned} \right.
\end{equation*}

\subsection{Analytic approximations of $\rho(\lambda)$}
Our next purpose is to deform the contour, but we need to discuss the decomposition
of $\rho(\lambda)$. Take
\begin{equation}\label{AA-1}
L:\left\{\lambda=\lambda_{0}+\lambda_{0}\alpha e^{\frac{3\pi i}{4}}:-\infty<\alpha\leq\sqrt{2}\right\}
\bigcup\left\{\lambda=-\lambda_{0}+\lambda_{0}\alpha e^{\frac{\pi i}{4}}:-\infty<\alpha\leq\sqrt{2}\right\},
\end{equation}
and
\begin{equation}\label{AA-2}
L_{\epsilon}:\left\{\lambda=\lambda_{0}+\lambda_{0}\alpha e^{\frac{3\pi i}{4}}:\epsilon<\alpha\leq\sqrt{2}\right\}
\bigcup\left\{\lambda=-\lambda_{0}+\lambda_{0}\alpha e^{\frac{\pi i}{4}}:\epsilon<\alpha\leq\sqrt{2}\right\},
\end{equation}
where $0<\epsilon\leq\sqrt{2}$.
\\
\noindent
\textbf{Lemma 3.} \cite{PD-1993}
As $0<\lambda_{0}\leq C$, there exists decomposition for the function $\rho(\lambda)$
\begin{equation}\label{AA-3}
\rho(\lambda)=h_{1}(\lambda)+h_{2}(\lambda)+R(\lambda),~~\lambda\in\mathbb{R},
\end{equation}
where $R(\lambda)$ is analytic in the complex plane and $h_{2}(\lambda)$ is analytically and continuously extended
to $L$. Additionally, $R(\lambda)$, $h_{1}(\lambda)$ and $h_{2}(\lambda)$ satisfy
\begin{equation}\label{AA-4}
\left\{ \begin{aligned}
&\left|e^{2it\theta(\lambda)}h_{1}(\lambda)\right|\lesssim t^{-l},~~\lambda\in\mathbb{R},\\
&\left|e^{2it\theta(\lambda)}h_{2}(\lambda)\right|\lesssim t^{-l},~~\lambda\in L,\\
&\left|e^{2it\theta(\lambda)}R(\lambda)(\lambda)\right|\lesssim e^{-16\epsilon^2\lambda_{0}^3},~~\lambda\in L_{\epsilon},
      \end{aligned} \right.
\end{equation}
where positive integer $l$ is free. It follows from the Schwartz conjugate representation of \eqref{AA-3} that
\begin{equation}\label{AA-5}
\rho^{\dag}(\bar{\lambda})=h_{1}^{\dag}(\bar{\lambda})+h_{2}^{\dag}(\bar{\lambda})+R^{\dag}(\bar{\lambda}),
\end{equation}
we can obtain the similar estimates for $e^{-2it\theta(\lambda)}h_{1}^{\dag}(\bar{\lambda})$,
$e^{-2it\theta(\lambda)}h_{2}^{\dag}(\bar{\lambda})$ and $e^{-2it\theta(\lambda)}R^{\dag}(\bar{\lambda})$
on the contour $\mathbb{R}\cup \bar{L}$.

$~~~~~~~~~~~~~~~~~~~~~~~~~~~$
{\rotatebox{0}{\includegraphics[width=8.0cm,height=6.5cm,angle=0]{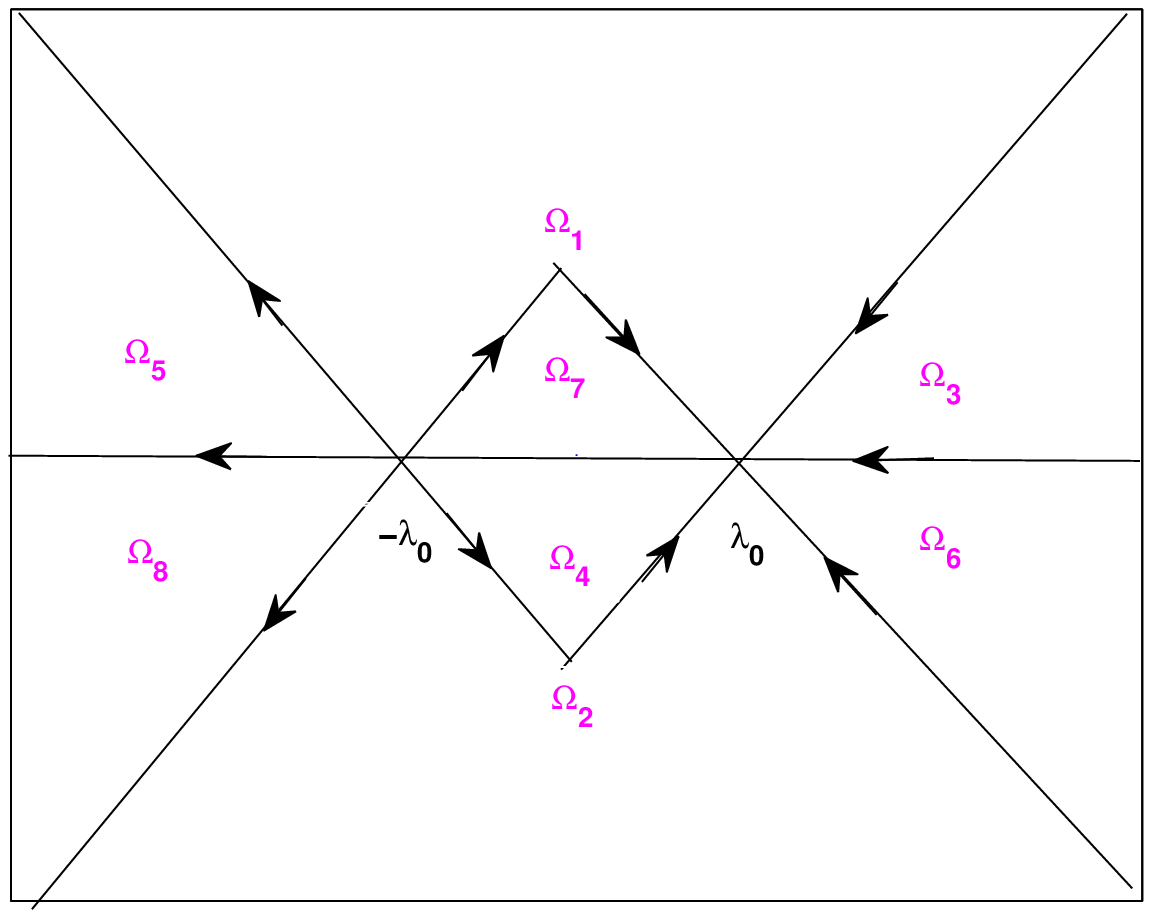}}}\\
\noindent {\small \textbf{Figure 2.}  The oriented jump contour $\Sigma$. \\}

\subsection{Contour deformation}
We rewrite $J^{\Delta}(x,t;\lambda)$ as $J^{\Delta}=(b_{-})^{-1}b_{+}$,
where $b_{\pm}=\mathcal {I}+\omega_{\pm}$, $\omega_{\pm}=\omega_{\pm}^{o}+\omega_{\pm}^{a}$,
\begin{align*}\label{CD-1}
b_{+}&=b_{+}^{o}b_{+}^{a}=\left(\mathcal {I}+\omega_{+}^{o}\right)\left(\mathcal {I}+\omega_{+}^{a}\right)\notag\\
&\triangleq
\left(
  \begin{array}{cc}
    \mathcal {I} & \det\delta(\lambda)e^{2i\theta}\delta(\lambda)h_{1}(\lambda) \\
    0 & 1 \\
  \end{array}
\right)\left(
         \begin{array}{cc}
           \mathcal {I} & \det\delta(\lambda)e^{2it\theta}\delta(\lambda)\left(h_{2}(\lambda)+R(\lambda)\right) \\
           0 & 1 \\
         \end{array}
       \right),\notag\\
b_{-}&=b_{-}^{o}b_{-}^{a}=\left(\mathcal {I}-\omega_{-}^{o}\right)\left(\mathcal {I}-\omega_{-}^{a}\right)\notag\\
&\triangleq
\left(
  \begin{array}{cc}
    \mathcal {I} & 0 \\
    -\frac{e^{-2it\theta}h_{1}^{\dag}(\bar{\lambda})\delta(\lambda)}{\det\delta(\lambda)} & 1 \\
  \end{array}
\right)\left(
         \begin{array}{cc}
           \mathcal {I} & 0 \\
           -\frac{e^{-2it\theta}\left(h_{2}^{\dag}(\bar{\lambda})+R^{\dag}(\bar{\lambda})\right)\delta^{-1}(\lambda)}{\det\delta(\lambda)} & 1 \\
         \end{array}
       \right).
\end{align*}

\noindent
\textbf{Lemma 4.} Take
\begin{equation}\label{CD-2}
M^{\sharp}(\lambda)=
\left\{ \begin{aligned}
&M^{\Delta}(\lambda),~~~~~~~~~~~\lambda\in\Omega_{1}\cup\Omega_{2},\\
&M^{\Delta}(\lambda)(b^{a}_{-})^{-1},~~\lambda\in\Omega_{3}\cup\Omega_{4}\cup\Omega_{5},\\
&M^{\Delta}(\lambda)(b^{a}_{+})^{-1},~~\lambda\in\Omega_{6}\cup\Omega_{7}\cup\Omega_{8}.
      \end{aligned} \right.
\end{equation}
As a result, the function $M^{\sharp}(\lambda)$ admits the RHP on the contour $\Sigma=L\cup \bar{L}\cup\mathbb{R}$ displayed in Fig.2,
\begin{equation}\label{CD-3}
\left\{ \begin{aligned}
&M_{+}^{\sharp}(\lambda)=M_{-}^{\sharp}(\lambda)J^{\sharp}(\lambda),~~\lambda\in\Sigma,\\
&M^{\sharp}\rightarrow\mathcal {I},~~~~~~~~~~~~~~~~~~~~\lambda\rightarrow\infty,
      \end{aligned} \right.
\end{equation}
where
\begin{equation}\label{CD-4}
J^{\sharp}=\left(b_{-}^{\sharp}\right)^{-1}b_{+}^{\sharp}\triangleq
\left\{ \begin{aligned}
&\left(b_{-}^{o}\right)^{-1}b_{+}^{o},~~\lambda\in\mathbb{R},\\
&\mathcal {I}^{-1}b_{+}^{a},~~~~~~~\lambda\in L,\\
&\left(b_{-}^{a}\right)^{-1}\mathcal {I},~~~\lambda\in \bar{L}.
      \end{aligned} \right.
\end{equation}
The above RHP \eqref{CD-3} can be obtained (see \cite{RB-1984}). Take
\begin{equation*}\label{CD-5}
\omega^{\sharp}=\omega^{\sharp}_{-}+\omega^{\sharp}_{+},~~\omega^{\sharp}_{\pm}=\pm b_{\pm}^{\sharp}\mp\mathcal {I}.
\end{equation*}
In the following, denote the Cauchy operators $C_{\pm}$ for $\lambda\in\Sigma$ by
\begin{equation*}\label{CD-6}
C_{\pm}f(\lambda)=\frac{1}{2\pi i}\int_{\Sigma}\frac{f(\xi)}{\xi-\lambda_{\pm}}d\xi,
\end{equation*}
where $f\in\mathscr{L}^{2}(\Sigma)$. Define
\begin{equation}\label{CD-7}
C_{\omega^{\sharp}}f=C_{+}\left(f\omega_{-}^{\sharp}\right)+C_{-}\left(f\omega_{+}^{\sharp}\right).
\end{equation}

\noindent
\textbf{Theorem 5}. \cite{RB-1984} Assume $\mu^{\sharp}(x,t;\lambda)\in\mathscr{L}^{2}(\Sigma)+\mathscr{L}^{\infty}(\Sigma)$ satisfies
\begin{equation*}\label{CD-8}
\mu^{\sharp}=\mathcal {I}+C_{\omega^{\sharp}}\mu^{\sharp}.
\end{equation*}
Thus
\begin{equation*}\label{CD-9}
M^{\sharp}(\lambda)=\mathcal {I}+\frac{1}{2\pi i}\int_{\Sigma}\frac{\mu^{\sharp}(\xi)\omega^{\sharp}(\xi)}{\xi-\lambda}d\xi,
\end{equation*}
represents the solution of the RHP \eqref{CD-3}.\\

\noindent
\textbf{Theorem 6}. The solutions $(u(x,t), v(x,t))$ for the CSS equation  \eqref{SS} can be expressed by
\begin{align}\label{CD-10}
\mathcal {U}(x,t)&=\left(
         \begin{array}{c}
           \textbf{u} \\
           \textbf{v} \\
         \end{array}
       \right)=2i\lim_{\lambda\rightarrow\infty}\left(\lambda M^{\sharp}(\lambda)\right)_{12}\notag\\
       &=-\frac{1}{\pi}\left(\int_{\Sigma}\left(\mu^{\sharp}(\xi)\omega^{\sharp}(\xi)d\xi\right)d\xi\right)_{12}\notag\\
       &=-\frac{1}{\pi}\left(\left(\int_{\Sigma}(1-C_{\omega^{\sharp}})^{-1}\mathcal {I}\right)(\xi)\omega^{\sharp}(\xi)d\xi\right)_{12}.
\end{align}
\noindent
\textbf{Proof:}
The similar result is provided in \cite{PD-1993}.

$~~~~~~~~~~~~~~~~~~~~~~~~~~~$
{\rotatebox{0}{\includegraphics[width=8.0cm,height=6.5cm,angle=0]{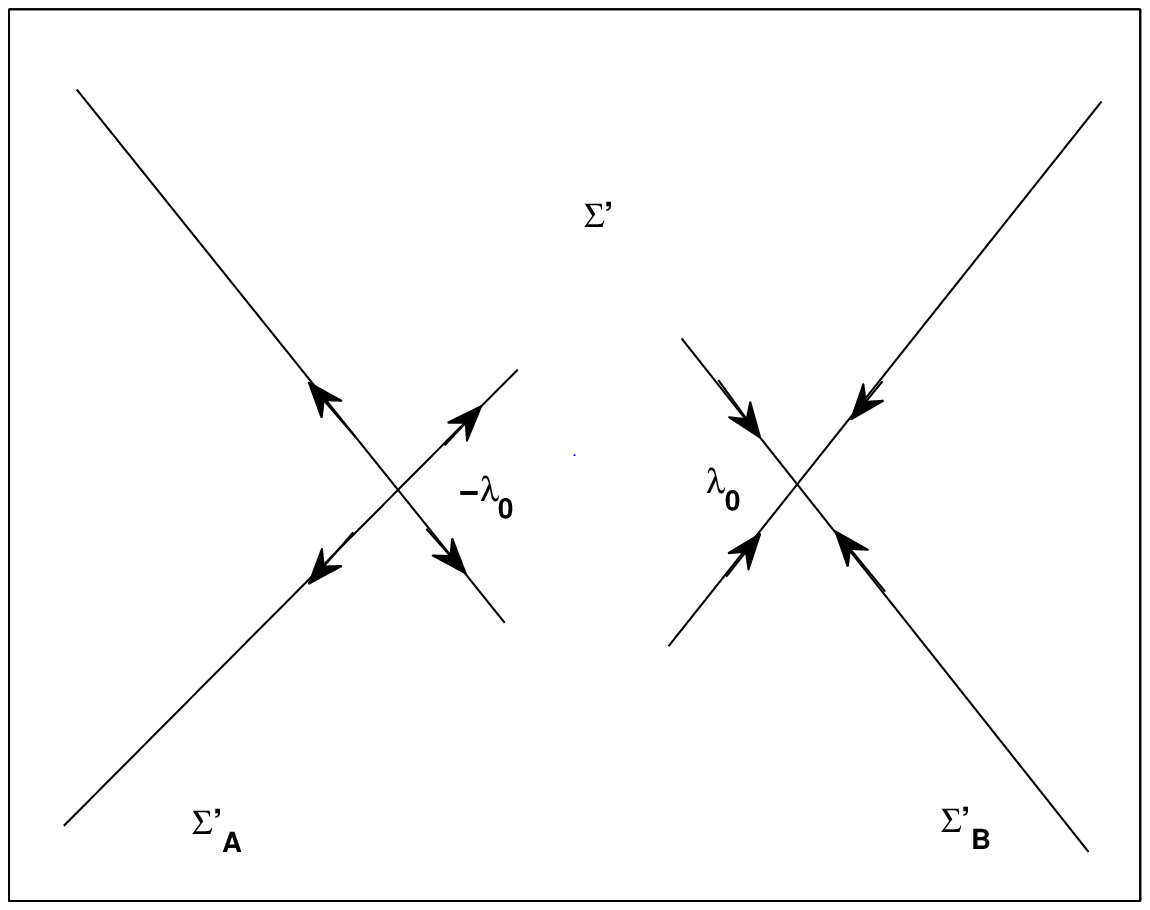}}}\\
\noindent {\small \textbf{Figure 3.}  The oriented contour $\Sigma'=\Sigma_{A}'\bigcup\Sigma_{B}'$.\\}

\subsection{Contour truncation}
As seen in Fig.3, take $\Sigma'=\Sigma(\mathbb{R}\cup L_{\epsilon}\cup \bar{L}_{\epsilon})$ with the orientation.
We plane to replace the RHP on $\Sigma$ with the truncated contour $\Sigma'$ by error control. Take
\begin{align*}\label{CT-1}
&\omega^{a}=\left\{ \begin{aligned}
&\omega^{\sharp},~~~~~\lambda\in\mathbb{R},\\
&\textbf{0},~~~~~\mbox{otherwise},
      \end{aligned} \right. \notag\\
&\omega^{b}=\left\{ \begin{aligned}
&\left(
  \begin{array}{cc}
    0 & \det\delta(\lambda)e^{2i\theta}\delta(\lambda) h_{2}(\lambda)\\
    0 & 0 \\
  \end{array}
\right),~~\lambda\in L,\\
&\left(
  \begin{array}{cc}
    0 & 0 \\
    \frac{e^{-2it\theta}h_{2}^{\dag}(\bar{\lambda})\delta^{-1}(\lambda)}{\det\delta(\lambda)} & 0 \\
  \end{array}
\right),~~~~~~~~~\lambda\in \bar{L},\\
&0,~~~~~~~~~~~~~~~~~~~~~~~~~~~~~~~~~~~~~~~~\mbox{otherwise},
      \end{aligned} \right.\notag\\
&\omega^{c}=\left\{ \begin{aligned}
&\left(
  \begin{array}{cc}
    0 & e^{2i\theta}\det\delta(\lambda)\delta(\lambda)R(\lambda)\\
    0 & 0 \\
  \end{array}
\right),~~\lambda\in L_{\epsilon},\\
&\left(
  \begin{array}{cc}
    0 & 0 \\
    \frac{e^{-2it\theta}R^{\dag}(\bar{\lambda})\delta^{-1}(\lambda)}{\det\delta(\lambda)} & 0 \\
  \end{array}
\right),~~~~~~~~~\lambda\in \bar{L}_{\epsilon},\\
&0,~~~~~~~~~~~~~~~~~~~~~~~~~~~~~~~~~~~~~~~~\mbox{otherwise}.
      \end{aligned} \right.
\end{align*}
Define $\omega'=\omega^{\sharp}-\omega^{a}-\omega^{b}-\omega^{c}$, so $\omega'=0$ on the contour $\Sigma\setminus\Sigma'$.
Thus, $\omega'$ is supposed on contour $\Sigma'$ and related to $R(\lambda)$ and $R^{\dag}(\bar{\lambda})$.

\noindent
\textbf{Lemma 7.} \cite{PD-1993}
For sufficiently small $\epsilon$, as $t\rightarrow\infty$
\begin{equation}\label{CT-2}
\left\{ \begin{aligned}
&\left\|\omega^{a}\right\|_{\mathscr{L}^{\infty}(\mathbb{R})\bigcap\mathscr{L}^{1}(\mathbb{R})}\lesssim t^{-l},\\\
&\left\|\omega^{b}\right\|_{\mathscr{L}^{\infty}(L\bigcup \bar{L})\bigcap\mathscr{L}^{1}(L\bigcup \bar{L})}\lesssim t^{-l},\\
&\left\|\omega^{c}\right\|_{\mathscr{L}^{\infty}(L_{\epsilon}\bigcup \bar{L}_{\epsilon})\bigcap\mathscr{L}^{1}(L_{\epsilon}\bigcup \bar{L}_{\epsilon})}\lesssim e^{-16\epsilon^2\tau},\\
&\|\omega'\|_{\mathscr{L}^{2}(\Sigma)}\lesssim\tau^{-1/4},~~\|\omega'\|_{\mathscr{L}^{1}(\Sigma)}\lesssim\tau^{-1/2},
      \end{aligned} \right.
\end{equation}
where $\tau=\lambda_{0}^3t$.

\noindent
%\textbf{Proof}.
%The proof of first three expressions in \eqref{CT-2}  follows from Proposition 3. Similar to \cite{}
%\begin{equation}\label{CT-3}
%|R(\lambda)|\lesssim\left(1+|\lambda|^5\right)^{-1},
%\end{equation}
%on the contour $\{\lambda=\lambda_{0}+\lambda_{0}\alpha e^{3\pi i/4}\}$.
%On this contour, using $\mbox{Re}i\theta(\lambda)\geq 8\lambda_{0}^3\alpha^2$
%we obtain
%\begin{equation}\label{CT-4}
%\left|\det(\delta(\lambda))e^{-2it\theta}\delta(\lambda)R(\lambda)\right|\lesssim e^{-16\lambda_{0}^2\alpha^2t}\left(1+|\lambda|^5\right)^{-1},
%\end{equation}
%After a direct calculation, we can obtain the last expression in \eqref{CT-2}.

\noindent
\textbf{Lemma 8.} \cite{PD-1993}
In the case $0<\lambda_{0}\leq C$, as $\tau\rightarrow\infty$,
the inverse $(1-C_{\omega'})^{-1}:\mathscr{L}^{2}(\Sigma)\rightarrow\mathscr{L}^{2}(\Sigma)$
exists, and has uniform boundedness
\begin{equation*}\label{CT-5}
\left\|\left(1-C_{\omega'}\right)^{-1}\right\|_{\mathscr{L}^{2}(\Sigma)}\lesssim 1.
\end{equation*}
Besides
\begin{equation*}\label{CT-5}
\left\|\left(1-C_{\omega^{\sharp}}\right)^{-1}\right\|_{\mathscr{L}^{2}(\Sigma)}\lesssim 1.
\end{equation*}

\noindent
\textbf{Lemma 9}. The integral equation has estimate as $\tau\rightarrow\infty$
\begin{equation}\label{CT-6}
\int_{\Sigma}\left(\left(1-C_{\omega^{\sharp}}\right)^{-1}\mathcal {I}\right)(\xi)\omega^{\sharp}(\xi)d\xi
=\int_{\Sigma}\left(\left(1-C_{\omega'}\right)^{-1}\mathcal {I}\right)(\xi)\omega'(\xi)d\xi+O\left(\frac{1}{\tau^{l}}\right).
\end{equation}

\noindent
\textbf{Proof:}
After a simple calculation, we find
\begin{align*}\label{L9-1}
\left(\left(1-\omega^{\sharp}\right)^{-1}\mathcal {I}\right)\omega^{\sharp}
&=((1-C\omega')^{-1}\mathcal {I})\omega'+\omega^{e}+((1-\omega')^{-1}\left(C_{\omega^{e}}\mathcal {I})\right)\omega^{\sharp}\notag\\
&+((1-\omega')^{-1}\left(C_{\omega'}\mathcal {I})\right)\omega^{e}\notag\\
&+\left((1-C_{\omega'})^{-1}C_{\omega^{e}}\left(1-C_{\omega^{\sharp}}\right)^{-1}\right)\left(C_{\omega^{\sharp}}\mathcal {I}\right)\omega^{\sharp}.
\end{align*}
The from Lemma 7, we have
\begin{equation*}\label{LP-2}
\left\{ \begin{aligned}
&\left\|\omega^{e}\right\|_{\mathscr{L}^{1}(\Sigma)}\leq\|\omega^{a}\|_{\mathscr{L}^{1}(\mathbb{R})}+\|\omega^{b}\|_{\mathscr{L}^{1}(L_{\epsilon}\cup \bar{L}_{\epsilon})}\lesssim \tau^{-l},\\
&\|\left(\left(1-C_{\omega'})^{-1}(C_{\omega^{e}}\mathcal {I}\right)\right)\omega^{\sharp}\|_{\mathscr{L}^{1}(\Sigma)}\leq
\|\left(1-C_{\omega'}\right)^{-1}\|_{\mathscr{L}^{2}(\Sigma)}\|C_{\omega^{e}}\mathcal {I}\|_{\mathscr{L}^{2}(\Sigma)}\|\omega^{\sharp}\|_{\mathscr{L}^{2}(\Sigma)}\\
&~~~~~~~~~~\leq\|\omega^{e}\|_{\mathscr{L}^{2}(\Sigma)}\|\omega^{\sharp}\|_{\mathscr{L}^{2}(\Sigma)}\lesssim t^{-l-1/4},\\
&\|\left(\left(1-C_{\omega'}\right)^{-1}(C_{\omega^{'}}\mathcal {I})\right)\omega^{e}\|_{\mathscr{L}^{1}(\Sigma)}\leq
\left\|\left(1-C_{\omega'}\right)^{-1}\right\|_{\mathscr{L}^{2}(\Sigma)}\|C_{\omega^{'}}\mathcal {I}\|_{\mathscr{L}^{2}(\Sigma)}\|\omega^{e}\|_{\mathscr{L}^{2}(\Sigma)}\\
&~~~~~~~~~~\leq\|\omega^{'}\left\|_{\mathscr{L}^{2}(\Sigma)}\right\|\omega^{e}\|_{\mathscr{L}^{2}(\Sigma)}\lesssim t^{-l-1/4},\\
&\|((1-C_{\omega'})^{-1}C_{\omega^{e}}(1-C_{\omega^{\sharp}})^{-1})(C_{\omega^{\sharp}}\mathcal {I})\omega^{\sharp}\|_{\mathscr{L}^{1}(\Sigma)}\\
&~~~~~~~~~~=\|(1-C_{\omega'})^{-1}\|_{\mathscr{L}^{2}(\Sigma)}\|C_{\omega^{e}}\|_{\mathscr{L}^{2}(\Sigma)}\|(1-C_{\omega^{\sharp}})^{-1}\|_{\mathscr{L}^{2}(\Sigma)}
\|C_{\omega^{\sharp}}\mathcal {I}\|_{\mathscr{L}^{2}(\Sigma)}\|\omega^{\sharp}\|_{\mathscr{L}^{2}(\Sigma)}\\
&~~~~~~~~~~\lesssim\|\omega^{e}\|_{\mathscr{L}^{\infty}(\Sigma)}\|\omega^{\sharp}\|^{2}_{\mathscr{L}^{2}(\Sigma)}\lesssim t^{-l-1/2}.
      \end{aligned} \right.
\end{equation*}
This finishes proof of Lemma 9.

\noindent
\textbf{Lemma 10}. The solution admits the following asymptotics, as $\tau\rightarrow\infty$
\begin{equation}\label{CT-7}
\mathcal {U}(x,t)=\left(
         \begin{array}{c}
           \textbf{u}(x,t) \\
           \textbf{v}(x,t) \\
         \end{array}
       \right)=-\frac{1}{\pi}\left(\int_{\Sigma'}\left(\left(1-C_{\omega'}\right)^{-1}\mathcal {I}\right)(x,t;\xi)\omega'(x,t,\xi)d\xi\right)_{12}+O\left(\frac{1}{\tau^{l}}\right).
\end{equation}
\textbf{Proof.} The lemma follows by Theorem 6 and Proposition 9.

Here take $L'=L\setminus L_{\epsilon}$ and $\Sigma'=L'\bigcup \bar{L}'$. Let $\mu'=(1-C_{\omega'})^{-1}\mathcal {I}$. Then
\begin{equation*}\label{CT-8}
M'(\lambda)=\mathcal {I}+\frac{1}{2\pi i}\int_{\Sigma'}\frac{\mu'(\xi)\omega'(\xi)}{\xi-\lambda}d\xi,
\end{equation*}
meets
\begin{equation*}\label{CT-9}
\left\{ \begin{aligned}
&M_{+}'(\lambda)=M_{-}'(\lambda)J'(\lambda),~~\lambda\in\Sigma',\\
&M'(\lambda)\rightarrow\mathcal {I},~~~~~~~~~~~~~~~~\lambda\rightarrow\infty,
      \end{aligned} \right.
\end{equation*}
where
\begin{align}\label{CT-10}
&J'=b_{-}'^{-1}b_{+}',~~b'_{-}=\mathcal {I},~~\lambda\in L',\notag\\
&b_{+}'=\left(
         \begin{array}{cc}
           \mathcal {I} & 0 \\
           e^{2it\theta(\lambda)}\det(\delta(\lambda))R(\lambda) & 1 \\
         \end{array}
       \right),\notag\\
&b'_{+}=\mathcal {I},~~b'_{-}=\left(
                               \begin{array}{cc}
                                 \mathcal {I} & -\frac{e^{-2it\theta(\lambda)}\delta^{-1}(\lambda)R^{\dag}(\bar{\lambda})}{\det \delta(\lambda)} \\
                                 0 & 1 \\
                               \end{array}
                             \right),~~\lambda\in \bar{L}'.
\end{align}

\subsection{Noninteraction of disconnected contour}
Choose $\omega'=\omega_{+}'+\omega_{-}'$, where $\omega'_{\pm}=\pm b'_{\pm}-\mp\mathcal {I}$.
Let the contour $\Sigma'=\Sigma_{\textbf{A}}'\bigcup\Sigma'_{\textbf{B}}$ and $\omega_{\pm}'=\omega'_{\textbf{A}_{\pm}}+\omega'_{\textbf{B}_{\pm}}$,
where $\omega'_{\textbf{A}_{\pm}}(\lambda)=0$ for $\lambda\in\Sigma'_{\textbf{B}}$, $\omega'_{\textbf{B}_{\pm}}(\lambda)$,
$\omega'_{\mathcal{\textbf{B}_{\pm}}}(\lambda)=0$ for $\lambda\in\Sigma'_{\textbf{A}}$.
Give the operators $C_{\omega'_{\textbf{A}}}$ and $C_{\omega'_{\textbf{B}}}$: $\mathscr{J}^{\infty}(\Sigma')+\mathscr{J}^{2}(\Sigma')\rightarrow\mathscr{J}^{2}(\Sigma')$
as in \eqref{CD-7}.

\noindent
\textbf{Lemma 11.} \cite{PD-1993}
\begin{align}\label{NDC-1}
&\|C_{\omega'_{\textbf{B}}C_{\omega'_{\textbf{A}}}}\|=\|C_{\omega_{\textbf{A}}'C_{\omega'_{\textbf{B}}}}\|_{\mathscr{J}^2(\Sigma')}\lesssim \lambda_{0}\tau^{-1/2},\notag\\
&\|C_{\omega'_{\textbf{B}}}C_{\omega'_{\textbf{A}}}\|_{\mathscr{J}^{\infty}(\Sigma')\rightarrow\mathscr{J}^{2}(\Sigma')},~~
\|C_{\omega'_{\textbf{A}}}C_{\omega'_{\textbf{B}}}\|_{\mathscr{J}^{\infty}(\Sigma')\rightarrow\mathscr{J}^{2}(\Sigma')}\leq\lambda_{0}\tau^{-3/4}.
\end{align}
\textbf{Proof:} Together with Lemma 7, Proposition 8 and 11, we can easily obtain \eqref{NDC-1}.

\noindent
\textbf{Lemma 12.} As $\tau\rightarrow\infty$
\begin{align}\label{NDC-2}
\int_{\Sigma'}\left(\left(1-C_{\omega'}\right)^{-1}\mathcal {I}\right)(\xi)\omega'(\xi)d\xi&=\notag\\
&\int_{\Sigma'_{\textbf{A}}}\left(\left(1-C_{\omega'_{\textbf{A}}}\right)^{-1}\mathcal {I}\right)(\xi)\omega'(\xi)d\xi\notag\\
&+\int_{\Sigma'_{\textbf{B}}}\left(\left(1-C_{\omega'_{\textbf{A}}}\right)^{-1}\mathcal {I}\right)(\xi)\omega'(\xi)d\xi+O\left(\frac{1}{\tau}\right).
\end{align}

\noindent
\textbf{Proof:} Form the following relation
\begin{align}\label{L12-1}
&\left(1-C_{\omega'_{\textbf{A}}}-C_{\omega'_{\textbf{B}}}\right)(1+C_{\omega'_{\textbf{A}}}\left(1-C_{\omega'_{\textbf{A}}}\right)^{-1}
+C_{\omega'_{\textbf{B}}}\left(1-C_{\omega'_{\textbf{B}}}\right)^{-1})\notag\\
&=1-C_{\omega'_{\textbf{B}}}C_{\omega'_{\textbf{A}}}\left(1-C_{\omega'_{\textbf{A}}}\right)^{-1}-
C_{\omega'_{\textbf{A}}}C_{\omega'_{\textbf{B}}}\left(1-C_{\omega'_{\textbf{B}}}\right)^{-1}
\end{align}
we find
\begin{align}\label{L12-22}
\left(1-C_{\omega'}\right)^{-1}=\left(1+C_{\omega'}\right)^{-1}&=1+
C_{\omega'_{\textbf{A}}}\left(1-C_{\omega'_{\textbf{A}}}\right)^{-1}+C_{\omega'_{\textbf{B}}}\left(1-C_{\omega'_{\textbf{B}}}\right)^{-1}\notag\\
&+\left(1+C_{\omega'_{\textbf{A}}}\left(1-C_{\omega'_{\textbf{A}}}\right)^{-1}+C_{\omega'_{\textbf{B}}}\left(1-C_{\omega'_{\textbf{B}}}\right)^{-1}\right)\notag\\
&\left(1-C_{\omega'_{\textbf{B}}}C_{\omega'_{\textbf{A}}}\left(1-C_{\omega'_{\textbf{A}}}\right)^{-1}
-C_{\omega'_{\textbf{A}}}C_{\omega'_{\textbf{B}}}\left(1-C_{\omega'_{\textbf{B}}}\right)^{-1}\right)^{-1}\notag\\
&\left(C_{\omega'_{\textbf{B}}}C_{\omega'_{\textbf{A}}}(1-C_{\omega'_{\textbf{A}}})^{-1}
+C_{\omega'_{\textbf{A}}}C_{\omega'_{\textbf{B}}}(1-C_{\omega'_{\textbf{B}}})^{-1}\right),
\end{align}
Then it follows from Lemma 7, Proposition 8 and 11 that Lemma 12.

\noindent
\textbf{Note.} We also write the restriction $C_{\omega'_{\textbf{A}}}|_{\mathscr{J}^{2}(\Sigma_{\textbf{A}}')}$ as $C_{\omega'_{\textbf{B}}}$.

\noindent
\textbf{Lemma 13.} As $\tau\rightarrow\infty$
\begin{align}\label{DNC-3}
\mathcal {U}(x,t)=\left(
         \begin{array}{c}
           \textbf{u} \\
           \textbf{v} \\
         \end{array}
       \right)&=-\frac{1}{\pi}\left(\int_{\Sigma'_{\textbf{A}}}\left(\left(1-C_{\omega'_{\textbf{A}}}\right)^{-1}\mathcal {I}\right)(x,t;\xi)\omega'_{\textbf{A}}(x,t,\xi)d\xi\right)_{12}\notag\\
&-\frac{1}{\pi}\left(\int_{\Sigma'_{\textbf{B}}}\left(\left(1-C_{\omega'_{\textbf{B}}}\right)^{-1}\mathcal {I}\right)(x,t;\xi)\omega'_{\textbf{B}}(x,t,\xi)d\xi\right)_{12}+O\left(\frac{1}{\tau}\right).
\end{align}

$~~~~~~~~~~~~~~~~~~~~~~~~~~~$
{\rotatebox{0}{\includegraphics[width=8.0cm,height=6.5cm,angle=0]{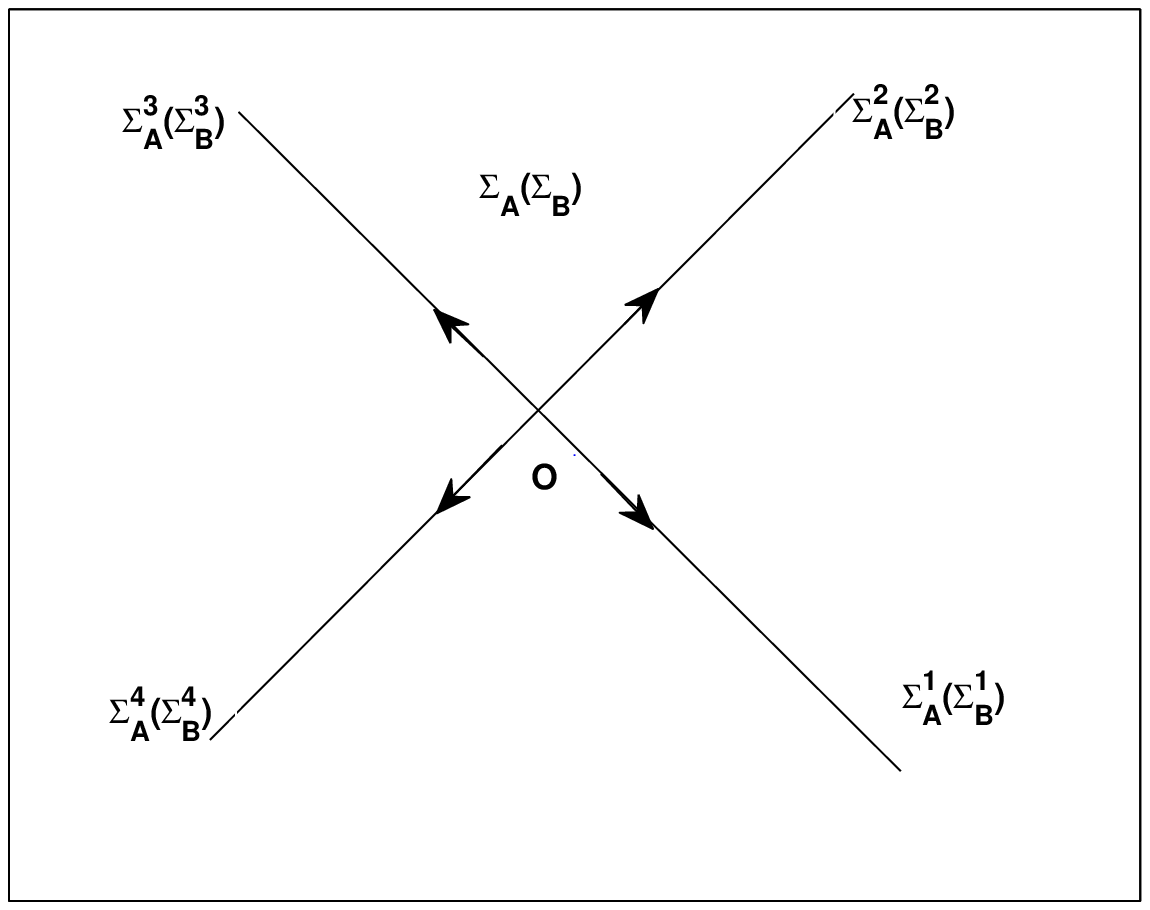}}}\\
\noindent {\small \textbf{Figure 4.}  The oriented contour $\Sigma_{A}$ or $\Sigma_{B}$.\\}

\subsection{The model Riemann-Hilbert Problem}

Extend $\Sigma'_{\textbf{A}}$ and $\Sigma_{\textbf{B}}'$ to the following contours
\begin{equation}\label{MRHP-1}
\left\{ \begin{aligned}
&\hat{\Sigma}_{\textbf{A}}'=\left\{\lambda=-\lambda_{0}+\lambda_{0}\alpha e^{\frac{\pm i\pi}{4}}: \alpha\in\mathbb{R}\right\},\\
&\hat{\Sigma}_{\textbf{B}}'=\left\{\lambda=-\lambda_{0}+\lambda_{0}\alpha e^{\frac{\pm 3i\pi}{4}}: \alpha\in\mathbb{R}\right\},
     \end{aligned} \right.
\end{equation}
respectively,
and give $\hat{\omega}'_{\textbf{A}}$, $\hat{\omega}'_{\textbf{B}}$ on $\hat{\Sigma}_{\textbf{A}}'$, $\hat{\Sigma}_{\textbf{B}}'$ as
\begin{align}\label{MRHP-2}
&\hat{\omega}_{\textbf{A}_{\pm}}'=\left\{ \begin{aligned}
&\omega_{A_{\pm}}'(\lambda),~~\lambda\in\Sigma'_{\textbf{A}},\\
&0,~~~~~~~~~~\lambda\in\hat{\Sigma}'_{\textbf{A}}\setminus\in\Sigma'_{\textbf{A}},
     \end{aligned} \right.\notag\\
&\hat{\omega}_{\textbf{B}_{\pm}}'=\left\{ \begin{aligned}
&\omega_{\textbf{B}_{\pm}}'(\lambda),~~\lambda\in\Sigma'_{\textbf{B}},\\
&0,~~~~~~~~~~\lambda\in\hat{\Sigma}'_{\textbf{B}}\setminus\in\Sigma'_{\textbf{B}},
     \end{aligned} \right.
\end{align}
Let $\Sigma_{\textbf{A}}$ and $\Sigma_{\textbf{B}}$ denote the contours $\{\lambda=\lambda_{0}\alpha e^{\pm\frac{i\pi}{4}}:\alpha\in\mathbb{R}\}$ shown in Fig.4.
The scaling operators $N_{\textbf{A}}$ and $N_{\textbf{B}}$ is given by
\begin{align}\label{MRHP-3}
N_{\textbf{A}}:&\mathscr{J}^2(\hat{\Sigma}'_{\textbf{A}})\rightarrow\mathscr{J}^2(\Sigma^{\textbf{A}})\notag\\
&f(\lambda)\mapsto (N_{\textbf{A}}f)(\lambda)=f\left(\frac{\lambda}{4\sqrt{3\lambda_{0}t}}-\lambda_{0}\right),\notag\\
N_{\textbf{B}}:&\mathscr{J}^2(\hat{\Sigma}'_{\textbf{B}})\rightarrow\mathscr{J}^2(\Sigma^{\textbf{B}})\notag\\
&f(\lambda)\mapsto (N_{\textbf{B}}f)(\lambda)=f\left(\frac{\lambda}{4\sqrt{3\lambda_{0}t}}+\lambda_{0}\right),
\end{align}
and take
\begin{equation}\label{MRHP-4}
\omega_{\textbf{A}}=N_{\textbf{A}}\hat{\omega}'_{\textbf{A}},~~\omega_{\textbf{B}}=N_{\textbf{B}}\hat{\omega}'_{\textbf{B}}.
\end{equation}
A simple replacement yields
\begin{equation}\label{MRHP-5}
C_{\hat{\omega}'_{\textbf{A}}}=N_{\textbf{A}}^{-1}C_{\omega_{\textbf{A}}}N_{\textbf{A}},
~~C_{\hat{\omega}'_{\textbf{B}}}=N_{\textbf{B}}^{-1}C_{\omega_{\textbf{A}}}N_{\textbf{B}},
\end{equation}
where $C_{\omega_{\textbf{A}}}:\mathscr{J}^{2}(\Sigma_{\textbf{A}})\Rightarrow\mathscr{J}^{2}(\Sigma_{\textbf{A}})$ is bounded, similar to $C_{\omega_{\textbf{B}}}$.

On the other hand
\begin{equation}\label{MRHP-6}
L_{\textbf{A}}=\left\{\lambda=\alpha\lambda_{0}4\sqrt{3\lambda_{0}t}e^{-\frac{3\pi i}{4}}:-\epsilon<\alpha<+\infty\right\},
\end{equation}
of $\Sigma_{\textbf{A}}$, we have
\begin{equation}\label{MRHP-7}
\omega_{\textbf{A}}=\omega_{A+}=\left(
                         \begin{array}{cc}
                           0 & \left(N_{\textbf{A}}s_{1}\right)(\lambda) \\
                           0 & 0 \\
                         \end{array}
                       \right),
\end{equation}
on $\bar{L}_{\textbf{A}}$ we obatin
\begin{equation}\label{mRHP-8}
\omega_{\textbf{A}}=\omega_{\textbf{A}-}=\left(
                         \begin{array}{cc}
                           0 & 0 \\
                           (N_{\textbf{A}}s_{2})(\lambda) & 0 \\
                         \end{array}
                       \right),
\end{equation}
where
\begin{align}\label{MRHP-9}
&s_{1}(\lambda)=\det \delta(\lambda) e^{2i\theta(\lambda)t}\delta(\lambda)R(\lambda),\notag\\
&s_{2}(\lambda)=\frac{e^{-2i\theta(\lambda)t}R^{\dag}(\bar{\lambda})\delta(\lambda)^{-1}}{\det\delta(\lambda)}.
\end{align}

\noindent
\textbf{Lemma 14.} As $\lambda\in L_{\textbf{A}}$, and $t\rightarrow\infty$
\begin{equation}\label{MRHP-10}
|(N_{\textbf{A}})\widetilde{\delta}(\lambda)|\lesssim t^{-l},
\end{equation}
where
\begin{equation}\label{MRHP-11}
\widetilde{\delta}(\lambda)=e^{2i\theta(\lambda)t}\left[\delta(\lambda)-\det \delta(\lambda)\mathcal {I}\right]R(\lambda).
\end{equation}

\noindent
\textbf{Proof:} It follows from \eqref{TRH-2}, \eqref{TRH-3} and \eqref{MRHP-11} that
\begin{equation}\label{L14-1}
\left\{ \begin{aligned}
&\widetilde{\delta}_{+}(\lambda)=e^{2it\theta}f(\lambda)+\widetilde{\delta}_{-}(\lambda)\left(1+|\gamma(\lambda)|^2\right),~~|\lambda|<\lambda_{0},\\
&\widetilde{\delta}(\lambda)\rightarrow0,~~\lambda\rightarrow\infty,
     \end{aligned} \right.
\end{equation}
where
\begin{equation*}\label{L14-2}
f(\lambda)=\delta_{-}\left(\gamma^{\dag}\gamma R-|\gamma|^2R\right)(\lambda).
\end{equation*}
By using Plemelj formula, the solution $\widetilde{\delta}(\lambda)$ reaches
\begin{equation*}\label{L14-3}
\widetilde{\delta}(\lambda)=X(\lambda)\int_{-\lambda_{0}}^{\lambda_{0}}\frac{e^{2it\theta(\xi) f(\xi)}}{X_{+}(\xi)(\xi-\lambda)}d\xi,
X(\lambda)=e^{\frac{1}{2\pi i}\int_{-\lambda_{0}}^{\lambda_{0}}\frac{1+|\gamma(\xi)|^2}{\xi-\lambda}d\xi},
\end{equation*}
Form
\begin{equation*}\label{L14-4}
\gamma^{\dag}\gamma R-|\gamma|^2R=\gamma^{\dag}\gamma(R-\rho)-|\gamma|^2(R-\rho),
\end{equation*}
we obtain $f(\lambda)=O(\lambda^2-\lambda_{0}^2)^{l}$. The we decompose $f(\lambda)$ two parts:
$f(\lambda)=f_{1}(\lambda)+f_{2}(\lambda)$, where $f_{2}(\lambda)$ is analytically and continuously
extended to $L_{t}$ and meets
\begin{equation}\label{L14-5}
\left\{ \begin{aligned}
&\left|e^{2it\theta(\lambda)}f_{1}(\lambda)\right|\lesssim\frac{1}{(1+|\lambda|^2)t^{l}},~~\lambda\in\mathbb{R},\\
&\left|e^{2it\theta(\lambda)}f_{1}(\lambda)\right|\lesssim\frac{1}{(1+|\lambda|^2)t^{l}},~~\lambda\in L_{t},
     \end{aligned} \right.
\end{equation}
in which
\begin{align*}\label{L14-6}
L_{t}:&\lambda=\left\{\lambda_{0}\alpha e^{3\pi i/4}:0\leq\alpha\leq\sqrt{2}-\frac{1}{\sqrt{2}t}\right\}\notag\\
&\bigcup\left\{\lambda=\lambda_{0}/t-\lambda_{0}+\lambda_{0}\alpha e^{\pi i/4}:0\leq\alpha\leq\sqrt{2}-\frac{1}{\sqrt{2}t}\right\},
\end{align*}
shown in Fig.5.

When $\lambda\in L_{\textbf{A}}$, we have
\begin{equation*}\label{L14-7}
N_{A}\widetilde{\delta}(\lambda)=\mathcal {I}_{1}+\mathcal {I}_{2}+\mathcal {I}_{3},
\end{equation*}
with
\begin{equation}\label{L14-8}
\left\{ \begin{aligned}
&|\mathcal {I}_{1}|\lesssim\int_{-\lambda_{0}}^{\lambda_{0}/t-\lambda_{0}}\frac{|f(\xi)|}{\xi+\lambda_{0}-\lambda/4\sqrt{3t\lambda_{0}}}d\xi\leq
t^{-l}\log\left|1-\frac{4\lambda_{0}\sqrt{3\lambda_{0}}}{\lambda\sqrt{t}}\right|\lesssim t^{-l-1/2},\notag\\
&|\mathcal {I}_{2}|\lesssim\int_{\lambda_{0}/t-\lambda_{0}}^{\lambda_{0}}\frac{|e^{2i\theta}f_{1}(\xi)|}{\xi+\lambda_{0}-\lambda/4\sqrt{3t\lambda_{0}}}d\xi\leq
t^{-l}\frac{\sqrt{2}t}{\lambda_{0}}\left(2\lambda_{0}-\lambda_{0}/t\right)\lesssim t^{-l+1},
     \end{aligned} \right.
\end{equation}
 With the help of Cauchy's Theorem, the original integral interval $(\lambda_{0}/t-\lambda_{0},\lambda_{0})$ in $\mathcal {I}_{3}$ can be replaced by contour $L_{t}$. Following the similar way, $|\mathcal {I}_{3}|\lesssim t^{-l+1}$, As a result, we can easily obtain lemma 14.

Following a similar way,
\begin{equation}\label{MRHP-12}
\left|(N_{\textbf{A}}\hat{\delta})(\lambda)\right|\lesssim t^{-l},~~\lambda\in \bar{L}_{\textbf{A}},~~t\rightarrow\infty,
\end{equation}
where
\begin{equation}\label{MRHP-13}
\hat{\delta}(\lambda)=e^{-2i\theta(\lambda)t}R^{\dag}(\bar{\lambda})\left[\delta(\lambda)^{-1}-[\det \delta^{-1}(\lambda)]\mathcal {I}\right].
\end{equation}

Take $J^{\textbf{A}^{0}}=(\mathcal {I}-\omega_{\textbf{A}^{0}-})^{-1}(\mathcal {I}-\omega_{\textbf{A}^{0}+})$, where
\begin{equation}\label{MRHP-14}
\omega_{\textbf{A}^{0}}=\omega_{\textbf{A}^{0}+}=\left\{ \begin{aligned}
&\left(
                                 \begin{array}{cc}
                                   0 & -(\delta_{\textbf{A}})^2(-\lambda)^{2iv}e^{i\lambda^2/2}\gamma^{\dag}(-\lambda_{0}) \\
                                   0 & 0 \\
                                 \end{array}
                               \right),~~\lambda\in\Sigma_{\textbf{A}}^{4},\\
 &\left(
                                 \begin{array}{cc}
                                   0 & (\delta_{\textbf{A}})^2(-\lambda)^{2iv}e^{i\lambda^2/2}\frac{\gamma^{\dag}(-\lambda_{0})}{1+|\gamma(-\lambda_{0})|^2} \\
                                   0 & 0 \\
                                 \end{array}
                               \right),~~\lambda\in\Sigma_{\textbf{A}}^{2},
                                    \end{aligned} \right.
\end{equation}
and
\begin{equation}\label{MRHP-15}
\delta_{\textbf{A}}=e^{\chi(-\lambda_{0})-8i\tau}(192\tau)^{-iv/2},
\end{equation}
with
\begin{equation}\label{MRHP-16}
\omega_{\textbf{A}^{0}}=\omega_{\textbf{A}^{0}-}=
\left\{ \begin{aligned}
&\left(
                                 \begin{array}{cc}
                                   0 & 0 \\
                                   -(\delta_{\textbf{A}})^-2(-\lambda)^{-2iv}e^{-i\lambda^2/2}\gamma(-\lambda_{0}) & 0 \\
                                 \end{array}
                               \right),~~\lambda\in\Sigma_{\textbf{A}}^{3},\\
                               &\left(
                                 \begin{array}{cc}
                                   0 & 0 \\
                                   (\delta_{\textbf{A}})^2(-\lambda)^{-2iv}e^{-i\lambda^2/2}\frac{\gamma(-\lambda_{0})}{1+|\gamma(-\lambda_{0})|^2} & 0 \\
                                 \end{array}
                               \right),~~\lambda\in\Sigma_{\textbf{A}}^{1}.
                                    \end{aligned} \right.
\end{equation}
It follows from \eqref{MRHP-10} and \cite{PD-1993} that
\begin{equation}\label{MRHP-17}
\|\omega_{\textbf{A}}-\omega_{\textbf{A}^{0}}\|_{\mathscr{J}^{\infty}}(\Sigma_{\textbf{A}})\bigcap\mathscr{J}^{1}(\Sigma_{\textbf{A}})\bigcap
\mathscr{J}^{2}(\Sigma_{\textbf{A}})\lesssim\lambda_{0}t^{1/2}\log(t).
\end{equation}
Therefore
\begin{align}\label{MRHP-18}
&\int_{\Sigma_{\textbf{A}}'}\left((1-C_{\omega'_{\textbf{A}}})^{-1}\mathcal {I}\right)(\xi)\omega'_{\textbf{A}}(\xi)d\xi\notag\\
&=\int_{\hat{\Sigma}_{\textbf{A}}}\left(\left(1-C_{\hat{\omega}'_{\textbf{A}}}\right)^{-1}\mathcal {I}\right)(\xi)\hat{\omega}'_{\textbf{A}}(\xi)d\xi\notag\\
&=\int_{\hat{\Sigma}'_{\textbf{A}}}(N_{\textbf{A}}^{-1}(1-C_{\omega_{\textbf{A}}})^{-1}N_{\textbf{A}}\mathcal {I})(\xi)\hat{\omega}'_{\textbf{A}}(\xi)d\xi\notag\\
&=\int_{\hat{\Sigma}'_{\textbf{A}}}\left(\left(1-C_{\omega_{\textbf{A}}}\right)^{-1}\mathcal {I}\right)\left(\xi+\lambda_{0}\right)4\sqrt{3t\lambda_{0}}N_{\textbf{A}}\hat{\omega}'_{\textbf{A}}
\left(\left(\xi+\lambda_{0}\right)4\sqrt{3t\lambda_{0}}\right)d\xi\notag\\
&=\frac{1}{4\sqrt{3t\lambda_{0}}}\int_{\Sigma_{\textbf{A}}}\left((1-C_{\omega_{\textbf{A}}})^{-1}\mathcal {I}\right)(\xi)\omega_{\textbf{A}}(\xi)\notag\\
&=\frac{1}{4\sqrt{3t\lambda_{0}}}\int_{\Sigma_{\textbf{A}}}\left((1-C_{\omega_{\textbf{A}^{0}}})^{-1}\mathcal {I}\right)(\xi)\omega_{\textbf{A}}(\xi)+O\left(\frac{\log t}{t}\right).
\end{align}
Together with a similar computation for \textbf{B} yields
\begin{align}\label{MRHP-19}
\mathcal {U}(x,t)=\left(
         \begin{array}{c}
           \textbf{u}(x,t) \\
           \textbf{v}(x,t) \\
         \end{array}
       \right)=&-\frac{1}{\pi}\frac{1}{4\sqrt{3\lambda_{0}t}}\left(\int_{\Sigma_{\textbf{A}}}((1-C_{\omega_{\textbf{A}^{0}}})^{-1}\mathcal {I})(\xi)\omega_{\textbf{A}^{0}}(\xi)d\xi\right)_{12}\notag\\
&-\frac{1}{\pi}\frac{1}{4\sqrt{3\lambda_{0}t}}\left(\int_{\Sigma_{\textbf{B}}}((1-C_{\omega_{\textbf{B}^{0}}})^{-1}\mathcal {I})(\xi)\omega_{\textbf{B}^{0}}(\xi)d\xi\right)_{12}
+O\left(\frac{\log t}{t}\right).
\end{align}

For $\lambda\in\mathbb{C}\setminus\Sigma_{A}$, let
\begin{equation}\label{MRHP-20}
M^{A^{0}}(\lambda)=\mathcal {I}+\frac{1}{2\pi i}\int_{\Sigma_{A}}\frac{((1-C_{\omega_{A^{0}}})^{-1})(\xi)\omega_{A^{0}}(\xi)}{\xi-\lambda}d\xi.
\end{equation}
Then $M^{\textbf{A}^{0}}$ admits
\begin{equation}\label{MRHP-21}
\left\{ \begin{aligned}
&M_{+}^{\textbf{A}^{0}}(\lambda)=M_{-}^{\textbf{A}^{0}}(\lambda)J^{\textbf{A}^{0}}(\lambda),~~\lambda\in\Sigma_{\textbf{A}},\\
&M^{\textbf{A}^{0}}(\lambda)\rightarrow\mathcal {I},~~~~~~~~~~~~~~~~~~~\lambda\rightarrow\infty.
         \end{aligned} \right.
\end{equation}
Particularly,
\begin{equation}\label{MRHP-22}
M^{\textbf{A}^{0}}(\lambda)=\mathcal {I}+\frac{M_{1}^{\textbf{A}^{0}}}{\lambda}+O\left(\frac{1}{\lambda^2}\right),~~\lambda\rightarrow\infty,
\end{equation}
then
\begin{equation}\label{MRHP-23}
M_{1}^{\textbf{A}^{0}}=-\frac{1}{2\pi i}\int_{\Sigma_{\textbf{A}}}\left(\left(1-C_{\omega_{\textbf{A}^{0}}}\right)^{-1}\mathcal {I}\right)(\xi)\omega_{\textbf{A}^{0}}(\xi)d\xi.
\end{equation}
A similar RHP for $\textbf{B}^{0}$ on $\Sigma_{\textbf{B}}$ reads
\begin{equation}\label{MRHP-24}
\left\{ \begin{aligned}
&M^{\textbf{B}^{0}}_{+}(\lambda)=M_{-}^{\textbf{B}^{0}}(\lambda)J^{\textbf{B}^{0}}(\lambda),~~\lambda\in\Sigma_{\textbf{B}},\\
&M^{\textbf{B}^{0}}(\lambda)\rightarrow\mathcal {I},~~~~~~~~~~~~~~~~~~~\lambda\rightarrow\infty.
         \end{aligned} \right.
\end{equation}
Utilizing \eqref{MRHP-14}-\eqref{MRHP-16} and $\omega_{\textbf{B}^{0}}$, one has
\begin{equation}\label{MRHP-25}
J^{\textbf{A}^{0}}(\lambda)=\bar{\tau}\left(J^{\textbf{B}^{0}}\right)(-\bar{\lambda})\tau.
\end{equation}
By uniqueness
\begin{equation*}\label{MRHP-26}
M_{1}^{\textbf{A}^{0}}(\lambda)=\bar{\tau}\left(M^{\textbf{B}^{0}}\right)(-\bar{\lambda})\tau,
\end{equation*}
and
\begin{equation*}\label{MRHP-27}
M^{\textbf{A}^{0}}_{1}=-\bar{\tau}\left(M_{1}^{\textbf{B}^{0}}\right)\tau.
\end{equation*}
Consequently, we have
\begin{equation}\label{MRHP-28}
\mathcal {U}(x,t)=\left(
         \begin{array}{c}
           \textbf{u}(x,t) \\
           \textbf{v}(x,t) \\
         \end{array}
       \right)=\frac{i}{\sqrt{12\lambda_{0}t}}\left(M_{1}^{\textbf{A}^{0}}-\sigma_{1}\left(\overline{M_{1}^{\textbf{A}^{0}}}\right)\right)_{12}+O\left(\frac{\log t}{t}\right).
\end{equation}

$~~~~~~~~~~~~~~~~~~~~~~~~~~~$
{\rotatebox{0}{\includegraphics[width=8.0cm,height=6.5cm,angle=0]{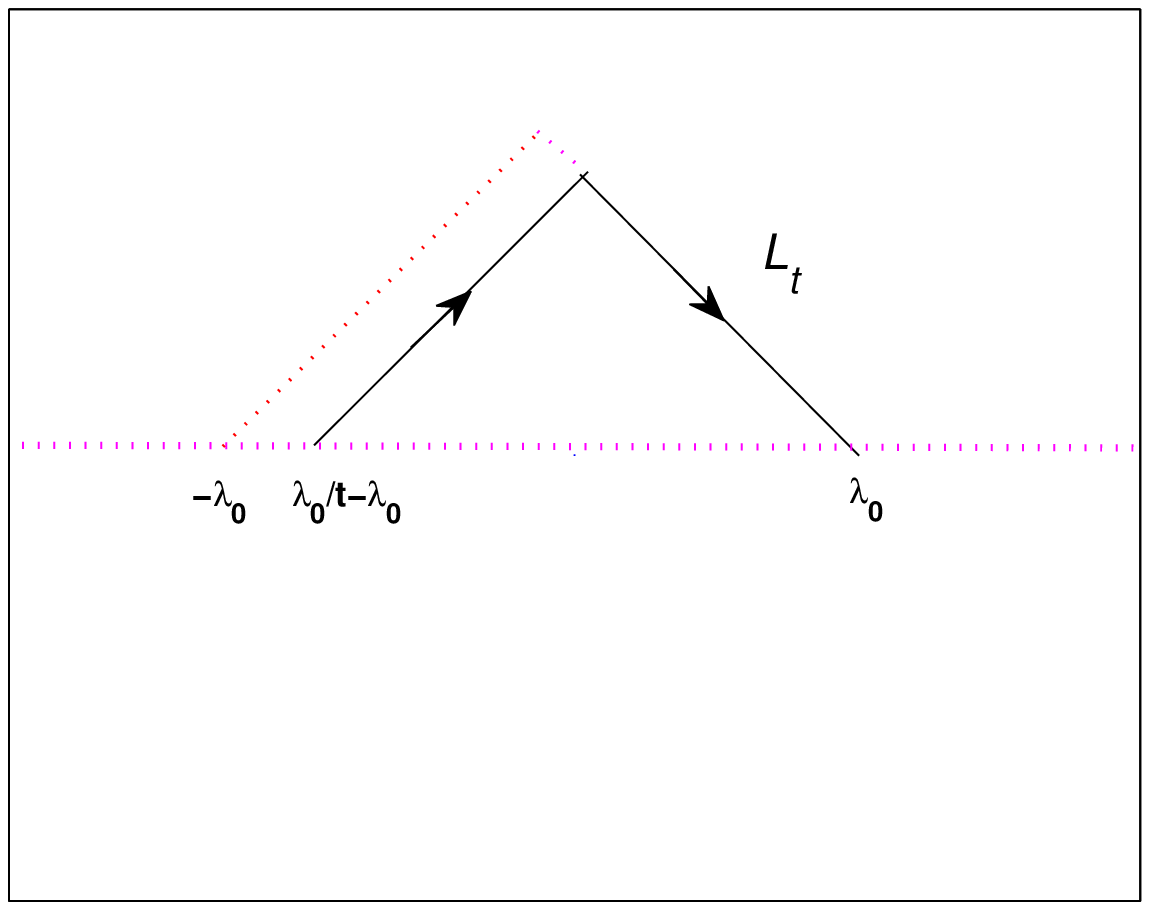}}}\\
\noindent {\small \textbf{Figure 5.}  The contour $L_{t}$.\\}

\subsection{The Final transformation}
To solve for $(M_{1}^{\textbf{A}^{0}})_{12}$ explicitly, the final transformation we set
\begin{equation*}\label{FTT-1}
\Psi(\lambda)=H(\lambda)(-\lambda)^{iv\sigma}e^{i\lambda^2\sigma/4},~~H(\lambda)=(\delta_{\textbf{A}})^{-\sigma}M^{\textbf{A}^{0}}(\lambda)(\delta_{\textbf{A}})^{\sigma}.
\end{equation*}
Therefore,
\begin{equation*}\label{FTT-2}
\Psi_{+}(\lambda)=\Psi_{-}(\lambda)v(-\lambda_{0}),~~v=(-\lambda)^{iv\hat{\sigma}}e^{-i\lambda^2\hat{\sigma}/4}(\delta_{\textbf{A}})^{-\hat{\sigma}}J^{\textbf{A}^{0}}.
\end{equation*}
Because of the jump matrix is independent of $\lambda$, we have
\begin{equation*}\label{FTT-3}
\frac{d\Psi_{+}(\lambda)}{d\lambda}=\frac{d\Psi_{-}(\lambda)}{dk}v(-\lambda_{0}).
\end{equation*}
Since $\frac{d\Psi(\lambda)}{d\lambda}$ and $\Psi$ have the same jump matrix along any of the rays,
it follows that $\frac{d\Psi(\lambda)}{d\lambda}\Psi^{-1}(\lambda)$ is holomorphic in the complex plane and  admits a polynomial dependence
on $\lambda$ at $\lambda\rightarrow\infty$. In reality
\begin{align*}\label{FTT-4}
&\frac{d\Psi(\lambda)}{d\lambda}\Psi^{-1}(\lambda)=\frac{i\lambda}{2}H(\lambda)\sigma H^{-1}(\lambda)-\frac{iv}{\lambda}H(\lambda)\sigma H^{-1}(\lambda)
+\frac{dH(\lambda)}{d\lambda}H^{-1}(\lambda)\notag\\
&=\frac{i\lambda}{2}\sigma-\frac{i}{2}\delta_{\textbf{A}}\sigma\left[\sigma,M_{1}^{\textbf{A}^{0}}\right]\delta_{\textbf{A}}^{-\sigma}+O\left(\frac{1}{\lambda}\right).
\end{align*}
Thus,
\begin{equation}\label{FTT-5}
\frac{d\Psi(\lambda)}{d\lambda}=\frac{i\lambda}{2}\sigma\Psi(\lambda)+\beta\Psi(\lambda),
\end{equation}
where
\begin{equation*}\label{FTT-6}
\beta=-\frac{i}{2}\delta_{\textbf{A}}^{\sigma}\left[\sigma,M_{1}^{\textbf{A}^{0}}\right]\delta_{\textbf{A}}^{-\sigma}=\left(
                                                                                  \begin{array}{cc}
                                                                                    0 & \beta_{12} \\
                                                                                    \beta_{21} & 0 \\
                                                                                  \end{array}
                                                                                \right).
\end{equation*}
Particularly
\begin{equation}\label{FTT-7}
(M_{1}^{\textbf{A}^{0}})_{12}=i(\delta_{\textbf{A}})^{2}\beta_{12}.
\end{equation}
Let
\begin{equation*}\label{FTT-8}
\Psi(\lambda)=\left(
                \begin{array}{cc}
                  \Psi_{11}(\lambda) & \Psi_{12}(\lambda) \\
                  \Psi_{21}(\lambda) & \Psi_{22}(\lambda) \\
                \end{array}
              \right).
\end{equation*}
It follows from \eqref{FTT-5} that
\begin{align}\label{FTT-9}
&\frac{d^2\beta_{12}\Psi_{11}(\lambda)}{d\lambda^2}=\left(\beta_{21}\beta_{12}+0.5i-\frac{\lambda^2}{2}\right)\beta_{21}\Psi_{11}(\lambda),\notag\\
&\frac{d^2\Psi_{22}(\lambda)}{d\lambda^2}=\left(\beta_{21}\beta_{12}-0.5i-\frac{\lambda^2}{4}\right)\Psi_{22}(\lambda),\notag\\
&\beta_{21}\Psi_{21}(\lambda)=\frac{1}{\beta_{12}}\left(\frac{d\beta_{21}\Psi_{11}(\lambda)}{d\lambda}-\frac{i\lambda}{2}\beta_{21}\Psi_{11}(\lambda)\right),\notag\\
&\beta_{21}\Psi_{12}(\lambda)=\frac{d\Psi_{22}(\lambda)}{d\lambda}+\frac{i\lambda}{2}\Psi_{22}(\lambda).
\end{align}
As everyone knows
\begin{equation*}\label{FTT-10}
g(\zeta)=c_{1}D_{a}(\zeta)+c_{2}D_{a}(-\zeta),
\end{equation*}
admits the solution of Weber's equation
\begin{equation*}\label{FTT-11}
g''(\zeta)+\left(a+\frac{1}{2}-\frac{\zeta^2}{4}\right)g(\zeta)=0,
\end{equation*}
where $D_{a}(\zeta)$ is the standard parabolic-cylinder function, and admits the following relations
\begin{align}\label{FTT-12}
&D'_{a}(\zeta)+\frac{\zeta}{2}D_{a}(\zeta)-aD_{a-1}(\zeta)=0,\notag\\
&D_{a}(\pm\zeta)=\Gamma(a+1)e^{i\pi a/2}D_{-a-1}(\pm i\zeta)+\frac{\Gamma e^{-i\pi a/2}}{\sqrt{2\pi}}D_{-1-a}(\mp i\zeta).
\end{align}
We know that as $\zeta\rightarrow\infty$ \cite{ET-1927}
\begin{align}\label{FTT-13}
&D_{a}(\zeta)=\notag\\
&\left\{ \begin{aligned}
&\zeta^{a}e^{-\zeta^2/4}\left(1+O\left(\frac{1}{\zeta^{2}}\right)\right),~~|\mbox{arg}\zeta|<\frac{3\pi}{4},\\
&\zeta^{a}e^{-\zeta^2/4}\left(1+O\left(\frac{1}{\zeta^{2}}\right)\right)-\frac{\sqrt{2\pi}}{\Gamma(-a)}e^{a\pi i+\zeta^2/4}\left(1+O\left(\frac{1}{\zeta^{2}}\right)\right),~~\frac{\pi}{4}<\mbox{arg}\zeta<\frac{5\pi}{4},\\
&\zeta^{a}e^{-\zeta^2/4}\left(1+O\left(\frac{1}{\zeta^{2}}\right)\right)-\frac{\sqrt{2\pi}}{\Gamma(-a)}e^{-a\pi i+\zeta^2/4}\left(1+O\left(\frac{1}{\zeta^{2}}\right)\right),~~-\frac{5\pi}{4}<\mbox{arg}\zeta<-\frac{\pi}{4},
      \end{aligned} \right.
\end{align}
where $\Gamma$ is the Gamma function.

Let $a=-i\beta_{21}\beta_{12}$, we have
\begin{equation*}\label{FTT-14}
\left\{ \begin{aligned}
&\beta_{21}\Psi_{11}(\lambda)=c_{1}D_{a}\left(e^{3\pi i/4}\lambda\right)+c_{2}D_{a}\left(e^{-\pi i/4}\lambda\right),\notag\\
&\Psi_{22}(\lambda)=c_{3}D_{-a}\left(e^{-3\pi i/4}\lambda\right)+c_{4}D_{-a}\left(e^{\pi i/4}\lambda\right).
      \end{aligned} \right.
\end{equation*}
As $\mbox{arg}\lambda\in(\frac{3\pi}{4},\pi)\bigcup(-\pi,-\frac{3\pi}{4})$ and $\lambda\rightarrow\infty$, we have
\begin{equation*}\label{FTT-15}
\Psi_{11}(\lambda)(-\lambda)^{-iv}e^{-i\lambda^2/4}\rightarrow\mathcal {I},~~
\Psi_{22}(\lambda)(-\lambda)^{-iv}e^{i\lambda^2/4}\rightarrow 1,
\end{equation*}
and
\begin{align}\label{FTT-16}
&\beta_{21}\Psi_{11}(\lambda)=\beta_{21}e^{-\pi v/4}D_{a}\left(e^{3\pi i/4}\lambda\right),~~v=-\beta_{21}\beta_{12},\notag\\
&\Psi_{22}(\lambda)=e^{-\pi v/4}D_{-a}\left(e^{-3\pi i/4}\lambda\right).
\end{align}
Thus
\begin{align}\label{FTT-17}
&\Psi_{21}(\lambda)=\frac{\beta_{21}e^{-\pi v/4}}{\beta_{21}\beta_{12}}\left(D_{a}'\left(e^{3\pi i/4}\lambda\right)-\frac{i\lambda}{2}D_{a}\left(e^{3\pi i/4}\lambda\right)\right),\notag\\
&=\beta_{21}e^{\pi(i+v)/4}D_{a-1}\left(e^{3\pi i/4}\lambda\right),\notag\\
&\Psi_{22}=e^{-\pi v/4}\left(D_{-a}'\left(e^{-3\pi i/4}\lambda\right)+\frac{i\lambda}{2}D_{-a}\left(e^{-3\pi i/4}\lambda\right)\right),\notag\\
&=ae^{\pi(i-v)/4}D_{-a-1}\left(e^{-3\pi i/4}\lambda\right).
\end{align}
As $\mbox{arg}\lambda\in(\frac{\pi}{4},\frac{3\pi}{4})$ and $\lambda\rightarrow\infty$, we have
\begin{equation*}\label{FTT-18}
\Psi_{11}(\lambda)(-\lambda)^{iv}e^{-i\lambda^2/4}\rightarrow\mathcal {I},~~\Psi_{22}(\lambda)(-\lambda)^{iv}e^{i\lambda^2/4}\rightarrow i,
\end{equation*}
then
\begin{equation*}\label{FTT-19}
\beta_{21}\Psi_{11}(\lambda)=\beta_{21}e^{3\pi v/4}D_{a}\left(e^{-\pi i/4}\lambda\right),~~
\Psi_{22}(\lambda)=e^{-\pi v/4}D_{-a}\left(e^{-3\pi i/4}\lambda\right).
\end{equation*}
So that
\begin{align}\label{FTT-20}
\Psi_{21}(\lambda)&=\frac{\beta_{21}e^{3\pi v/4}}{\beta_{21}\beta_{12}}\left(D'_{a}(e^{-\pi i/4}\lambda)
-\frac{i\lambda}{2}D_{a}\left(e^{-\pi i/4}\lambda\right)\right)\notag\\
&=\beta_{21}e^{-3\pi(i-v)/4}D_{a-1}\left(e^{-\pi i/4}\lambda\right),\notag\\
&\beta_{21}\Psi_{12}(\lambda)=ae^{\pi(i-v)/4}D_{-a-1}\left(e^{-3\pi i/4}\lambda\right).
\end{align}
Along the ray $\mbox{arg}\lambda=3\pi/4$, we have
\begin{align*}\label{FTT-21}
&\Psi_{+}(\lambda)=\Psi_{-}(\lambda)\left(
                                     \begin{array}{cc}
                                       \mathcal {I} & 0 \\
                                       -\gamma(-\lambda_{0}) & 1 \\
                                     \end{array}
                                   \right),\notag\\
&\beta_{21}e^{\pi(i-v)/4}D_{a-1}\left(e^{3\pi i/4}\lambda\right)=\beta_{21}e^{-3\pi(i-v)/4}D_{a-1}\left(e^{-\pi i/4}\lambda\right)-\gamma(-\lambda_{0})e^{-\pi v/4}D_{-a}\left(e^{-3\pi i/4}\lambda\right),\notag\\
&D_{-a}\left(e^{-3\pi i/4}\lambda\right)=\frac{\Gamma(-a+1)e^{-\pi ia/2}}{\sqrt{2\pi}}D_{a-1}\left(e^{-\pi i/4}\lambda\right)+\frac{\Gamma(-a+1)e^{\pi ia/2}}{\sqrt{2\pi}}D_{a-1}\left(e^{3\pi i/4}\lambda\right),\notag\\
&\beta_{21}=\Gamma(-a+1)e^{-\pi v/2}e^{3\pi i/4}\gamma(-\lambda_{0})=\frac{-v\Gamma(-iv)e^{\pi v/2}e^{-3\pi i/4}}{\sqrt{2\pi}}\gamma(-\lambda_{0}).
\end{align*}
It is clear to see that $\Psi^{-1}(\lambda)$ and $\Psi^{\dag}(\bar{\lambda})$ satisfy the same RHP. Due to the uniqueness, we get
\begin{equation}\label{FTT-22}
\Psi^{-1}(\lambda)=\Psi^{\dag}(\bar{\lambda},
\end{equation}
and thus
\begin{equation}\label{FTT-23}
\beta_{12}=-\beta_{21}^{\dag}=\frac{v\Gamma(iv)e^{\pi v/2}e^{-\pi i/4}}{\sqrt{2\pi}}\sigma_{1}\gamma^{T}(\lambda_{0}).
\end{equation}
It follows from $\beta_{21}\beta_{12}=-v$ and $\Gamma(-iv)=\bar{\Gamma}(iv)$ that
\begin{equation}\label{FTT-24}
\frac{v\Gamma(iv)e^{\pi v/2}}{\sqrt{2\pi}}=\frac{\sqrt{v}}{|\gamma(\lambda_{0})|}.
\end{equation}

Summarizing the above analysis, the following theorem holds.

\noindent
\textbf{Theorem 15} Let $(u_{0}$, $v_{0})$ belong to the Schwartz space $\mathcal {S}(\mathbb{R})$.
Then suppose $u(x,t)$, $v(x,t)$ can solve the CSS equation \eqref{SS}. As $x<0$ and $|\frac{x}{t}|$ is bounded,
the solutions $u(x,t)$, $v(x,t)$ admits the following leading asymptotics
\begin{equation*}\label{T15-1}
\left\{ \begin{aligned}
&u(x,t)=\frac{u_{as}(x,t)}{\sqrt{t}}+O\left(\frac{\log t}{t}\right),\\
&v(x,t)=\frac{v_{as}(x,t)}{\sqrt{t}}+O\left(\frac{\log t}{t}\right),
      \end{aligned} \right.
\end{equation*}
where
\begin{align*}\label{T15-2}
&u_{as}(x,t)=\frac{\sqrt{v}}{\sqrt{12\lambda_{0}}|\gamma(\lambda_{0})|}\left(|\gamma_{2}(\lambda_{0})|e^{i(\phi+\mbox{arg} \gamma_{2}(\lambda_{0}))}
+|\gamma_{1}(\lambda_{0})|e^{-i(\phi+\mbox{arg}\gamma_{1}(\lambda_{0}))}\right),\notag\\
&v_{as}(x,t)=\frac{\sqrt{v}}{\sqrt{12\lambda_{0}}|\gamma(\lambda_{0})|}\left(|\gamma_{4}(\lambda_{0})|e^{i(\phi+\mbox{arg} \gamma_{4}(\lambda_{0}))}
+|\gamma_{3}(\lambda_{0})|e^{-i(\phi+\mbox{arg}\gamma_{3}(\lambda_{0}))}\right),\notag\\
&\phi=16t\lambda_{0}^3+\mbox{arg}\Gamma(iv)+v\log(192\lambda_{0}^3t)+\frac{1}{\pi}
\int_{-\lambda_{0}}^{\lambda_{0}}\log\left(\frac{1+|\gamma(\xi)|^2}{1+|\gamma(\lambda_{0})|^2}\right)\frac{d\xi}{\xi+\lambda_{0}}-\frac{5\pi}{4},\notag\\
&v=\frac{1}{2\pi}\log\left(1+|\gamma(\lambda_{0}|^{2}\right),
\end{align*}
and $\gamma_{1}$, $\gamma_{2}$, $\gamma_{3}$, $\gamma_{4}$ is the 1,2,3,4-th component of the vector function $\gamma$ given by \eqref{RHH-9},
$\lambda_{0}=\sqrt{-x/(12t)}$, $\Gamma$ is a Gamma function.

\section{Further discussions}
In this work, we have obtained a $5\times5$ matrix Riemann-Hilbert problem to tackle the Cauchy problem for the CSS equation \eqref{SS} on the line,
which can help us to obtain a representation for the solution of the CSS equation \eqref{SS}.
We then employ the approach of the Deift-Zhou steepest descent to discuss the long-time asymptotics of the CSS equation \eqref{SS}.
Similarly to \cite{PRE-2018}, starting from the CSS equation \eqref{SS}, if we impose the solution constraint
\begin{equation}\label{FD-1}
v(x,t)=\bar{u}(-x,t),
\end{equation}
we can obtain
\begin{equation}\label{FD-2}
u_{t}+\left\{u_{xxx}+6\left(|u|^{2}+|u(-x,t)|^{2}\right)u_{x}+3u\left(|u|^{2}+|u(-x,t)|^{2}\right)_{x}\right\}=0,
\end{equation}
which is a nonlocal equation of reverse-time type.
If we impose the solution constraint
\begin{equation}\label{FD-3}
v(x,t)=\bar{u}(-x,-t),
\end{equation}
the coupled Sasa-Satsuma \eqref{SS} reduces to
\begin{equation}\label{FD-4}
u_{t}+\left\{u_{xxx}+6\left(|u|^{2}+|u(-x,-t)|^{2}\right)u_{x}+3u\left(|u|^{2}+|u(-x,-t)|^{2}\right)_{x}\right\}=0,
\end{equation}
which is a nonlocal equation of reverse-space-time type.
These two equations differ from the other nonlocal equations of reverse-time and reverse-space-time types \cite{MJA-2017,ZZN-2017}
in the nonlinear terms.
Both of our nonlocal equations \eqref{FD-2} and \eqref{FD-4} are also integrable which have clear physical meanings, and their Lax pairs are
\begin{equation*}
\left\{ \begin{aligned}
  &\varphi_{x}=i\lambda\sigma\varphi+\textbf{U}(x,t)\varphi,\\
  &\varphi_{t}=4i\lambda^{3}\sigma\varphi+\textbf{V}\varphi,
                        \end{aligned} \right.
\end{equation*}
where
\begin{align*}\label{SA-3}
\textbf{U}=\left(
             \begin{array}{ccccc}
               0 & 0 & 0 & 0 & u(x,t) \\
               0 & 0 & 0 & 0 & \bar{u}(x,t) \\
               0 & 0 & 0 & 0 & u(-x,t) \\
               0 & 0 & 0 & 0 & u(-x,t) \\
               -\bar{u}(x,t) & -u(x,t) & -u(-x,t) & -u(-x,t) & 0 \\
             \end{array}
           \right),
\end{align*}
and
\begin{align*}
\textbf{U}=\left(
             \begin{array}{ccccc}
               0 & 0 & 0 & 0 & u(x,t) \\
               0 & 0 & 0 & 0 & \bar{u}(x,t) \\
               0 & 0 & 0 & 0 & u(-x,-t) \\
               0 & 0 & 0 & 0 & u(-x,-t) \\
               -\bar{u}(x,t) & -u(x,t) & -u(-x,-t) & -u(-x,-t) & 0 \\
             \end{array}
           \right)
\end{align*}
with
\begin{equation*}\label{SA-4}
\textbf{V}(x,t,\lambda)=4\lambda^2\textbf{U}-2i\lambda\sigma\left(\textbf{U}_{x}-\textbf{U}^{2}\right)
+\left(\textbf{U}_{x}U-\textbf{U}\textbf{U}_{x}\right)-\textbf{U}_{xx}+2\textbf{U}^{3}.
\end{equation*}

Finally, we state that there exist several methods to derive exact solutions for the most NLS equations, such as
Darboux transformation, Inverse scattering transform,  RH approach, Deift-Zhou steepest descent method, Hirota method, dressing method,
Wronskian technique etc.
Consequently, it is very worthy to consider whether the nonlocal equations \eqref{FD-2} and \eqref{FD-4}  can be
solved by using these approaches? These will be left for future discussions.

\section*{Acknowledgements}
\hspace{0.3cm}This work is supported by the National Key Research and Development Program of
China under Grant No. 2017YFB0202901 and the National Natural Science Foundation of China under Grant No.11871180.

%\end{CJK*}
\end{document}